\documentclass[%
 reprint,
 amsmath,amssymb,
 aps,
]{revtex4-2}

\usepackage{graphicx}
\usepackage{dcolumn}
\usepackage{bm}
\usepackage{color}
\usepackage{upgreek}
\usepackage{latexsym}
\usepackage{graphicx}
\usepackage{epstopdf}
\usepackage{color}
\usepackage[colorlinks,
            linkcolor=blue,
            anchorcolor=blue,
            citecolor=blue 
            ]{hyperref}

\begin{document}

\preprint{APS/123-QED}

\title{Laser field strength dependence of solid high-harmonic generation from doped systems}

\author{Tian-Jiao Shao}
\email[]{shaotj@nit.zju.edu.cn}
\affiliation{School of Information Science and Engineering, NingboTech University, Ningbo 315100, China}
\affiliation{State Key Laboratory of Magnetic Resonance and Atomic and Molecular Physics,
Wuhan Institute of Physics and Mathematics, Innovation Academy for Precision Measurement Science and Technology, Chinese Academy of Sciences, Wuhan 430071, China}


\date{\today}

\begin{abstract}
We have theoretically studied the field-strength dependent high-harmonic generation (HHG) in doped systems like nano-size or bulk materials. Our results show when the amplitude of the vector potential $A_{peak}$ of the driving laser reaches the half-width of the Brillouin zone ($\pi$/$a_{0}$), the harmonic yield of the undoped systems is larger than the doped systems. The band-climbing process enhances the interband transition of HHG for higher conduction bands. When $A_{peak}$ is below $\pi$/$a_{0}$, the harmonic yield of the doped systems is stronger than the undoped systems. The atomic doping density also influences the field-strength dependent spectra. 

\end{abstract}

\maketitle


\section{Introduction}

With the development of ultra-fast laser technology, high-order harmonic generation (HHG) in gas-medium was first observed experimentally in 1987 \cite{mcpherson1987studies}. The three-step model was proposed to explain the physical mechanism of HHG in gas \cite{corkum1993plasma,lewenstein1994theory}. Based on the HHG, isolated attosecond pulses synthesis in experiments \cite{paul2001observation,hentschel2001attosecond}, characterization of electronic structure and ultrafast dynamics are realized \cite{lein2002electron,itatani2004tomographic,probing2014bian}. With the development of mid-infrared laser \cite{dubietis1992powerful, ishii2014carrier}, the experiment of HHG in solids has been carried out and has become a research hotspot. The density and highly oriented arrangement of ions in solid make the solid HHG have many differences from its counterpart in gas, such as the double-plateau structure of HHG spectra \cite{ndabashimiye2016solid, you2017laser}, unique dependence on the ellipticity of driving field \cite{ghimire2011observation,yoshikawa2017high, tancogne2017ellipticity}, linear dependence on the driving field strength \cite{ghimire2011observation,Vampa2014,vampa2015semiclassical}, and so forth.  

In addition, the electronic structure and band energies of solid targets can be designed through means of material processing technology, and then be used to control the HHG process. Unlike the gas media, solid target materials can grow nanostructures on the surface \cite{han2016high,vampa2017plasmon,liu2018enhanced}, modification \cite{sivis2017tailored}, reduce the dimension of material \cite{lou2020ellipticity,liu2017high,mcdonald2017enhancing}, stacking \cite{le2018high}, heterostructure \cite{alonso2021giant}, apply stress and strain \cite{qin2018strain,strain2019shao,tamaya2022shear}, doping \cite{nefedova2021enhanced,wang2017roles,huang2017high,yu2019enhanced,high2019jia,mrudul2020high,pattanayak2020influence,zhao2021impact} and other material engineering methods. These methods can change the electronic structure of the solid target, form the surface plasmon polaritons \cite{han2016high,vampa2017plasmon,liu2018enhanced}, change the local solid medium and driving field \cite{sivis2017tailored}, and then HHG can be enhanced. 

The acceptor dopant will form an impurity energy level that is unoccupied between the bandgap. Driven by the electric field, the electrons in the valence band (VB) of the acceptor-doped semiconductor are easier to excite to impurity energy level and more holes can be created in VB compared with the undoped system. While for the donor-doped semiconductor, the electrons in the occupied impurity energy level are easier to excite to the conduction band (CB) compared with other states in VB, and more electrons can be created in the CB compared with the undoped semiconductor. 

The physical scheme of the HHG in doped systems has been theoretically investigated \cite{huang2017high,yu2019enhanced,high2019jia,mrudul2020high,pattanayak2020influence,zhao2021impact}. The addition of impurity energy level plays the role of “ladder” for the optical transition from the band with lower energy to higher energy and greatly affects the HHG dynamics \cite{huang2017high,yu2019enhanced,zhao2021impact}. In 2017, Huang $et. al$ theoretically found the second plateau of the HHG spectra of donor-doped semiconductors is enhanced compared with the undoped system by using the time-dependent Schrödinger equation (TDSE) method \cite{huang2017high}. Their work indicates the narrower bandgap and Brillouin zone in the donor-doped system strengthen the population in the CBs and improves the HHG emission. In 2019, Yu $et. al$ report the enhancement of the HHG in the donor-doped semiconductor by using time-dependent density functional theory (TDDFT) simulation. Their work shows that atomic-like impurity-state can be explained by a semi-classically three-step model \cite{yu2019enhanced}. In 2019, Jia $et. al$ investigated HHG magnetically-doped topological insulators Bi$_{2}$Se$_{3}$ and found the crucial interplay between laser polarization and the symmetry of material \cite{high2019jia}. In 2020, Mrudul $et. al$ investigated the spin-polarised defects in two-dimensional hexagonal boron nitride by using TDDFT \cite{mrudul2020high}. Their calculation revealed that different spin channels are influenced differently by the spin-polarised defect. In the same year, Pattanayak $et. al$'s work point out the impact of vacancy defects on the cutoff and yield of HHG \cite{pattanayak2020influence}. In 2021, Zhao $et. al$ investigated the influence of the donor- and acceptor-doped impurities on HHG and found that the impurity energy level in the middle of the bandgap will lead to a higher yield of HHG emission \cite{zhao2021impact}. In 2021, V. E. Nefedova $et. al$ experimentally investigated HHG in Cr-doped MgO \cite{nefedova2021enhanced}. An enhancement of the HHG is found even though the defect concentration is low which accords with theoretical prediction \cite{huang2017high,yu2019enhanced,zhao2021impact}. 

However, the field strength-dependent HHG from the doped systems hasn't been investigated, especially for large driving field strength. In this work, we found that the enhancement of HHG in the donor-doped system is limited to a particular range of the field strength of the driving laser. The enhancement by doping is no longer preserved when the amplitude of the vector potential $A_{peak}$ is around or above the half-width $\pi$/$a_{0}$ of the Brillouin zone with $a_{0}$ being the lattice constant.

\begin{figure*}[t]
\includegraphics[width=16.0cm,angle=0]{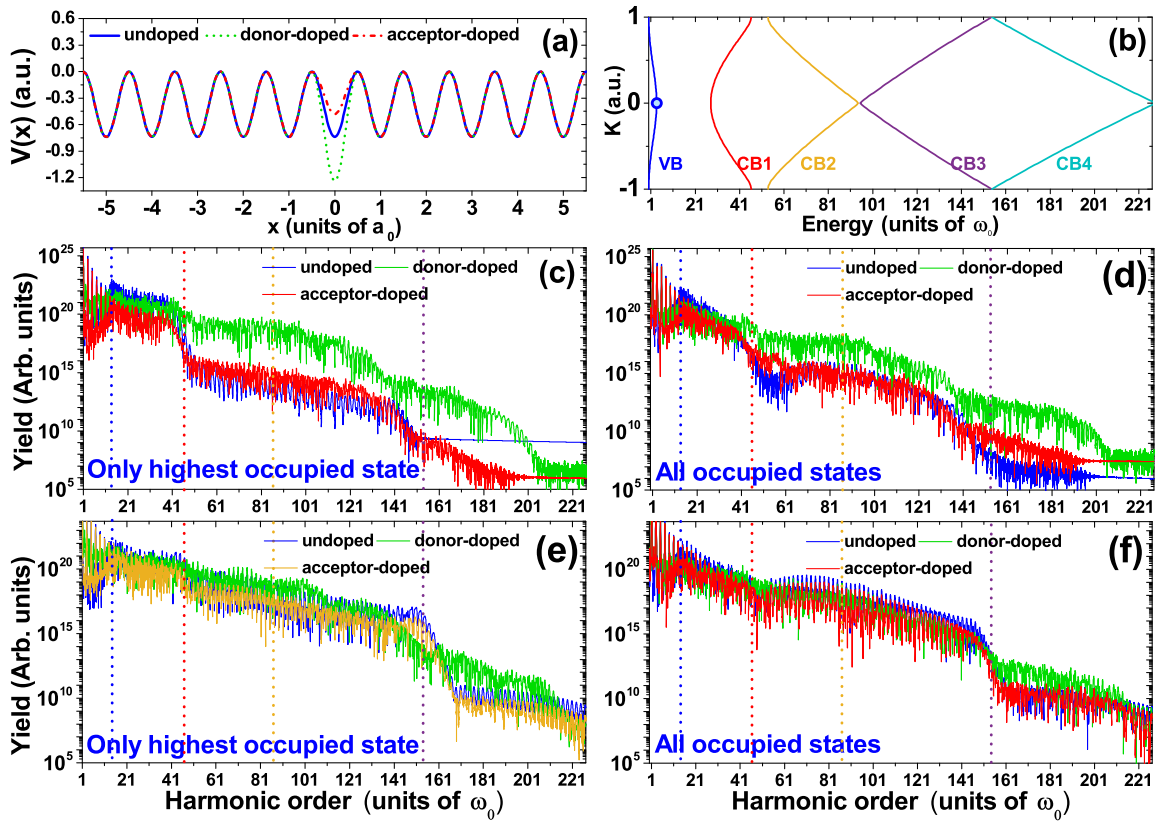}
\caption{\label{fig1} (a) The blue solid line, green dashed line, and red dashed-dotted line show the potential of the undoped, donor-doped, and acceptor-doped systems, respectively. (b) The band structure of the undoped system was calculated by using the Bloch states expansion method. The blue scatter in the center of VB is the highest occupied state for the undoped system which contributes most to the HHG than the other occupied states. (c, d) Comparison of the HHG spectra in the undoped system, donor-doped, and acceptor-doped system driven by eight-cycle Gaussian laser pulse with field strength $E_{0}$=0.0038 a.u. and a central wavelength of 4000 nm. (e, f) Same as the middle row except for the field strength changes to $E_{0}$=0.0045 a.u.}
\end{figure*}

\section{Methods}

The calculations in this work are on the basis of the solution of the TDSE in periodic potential \cite{wu2015high,guan2016high,Wang2021Model}. The Mathieu-type potential is used to describe the periodic potential. For undoped periodic lattice, the potential is given by, $V(x)=-V_{0}[1+cos(2\pi x/a_{0})]$ with $V_{0}$=0.37 a.u and the lattice constant $a_{0}$=8 a.u. Mathieu-type potentials are often used to simulate HHG in solids. The periodic potential of the undoped system is shown by the blue solid line in Fig. \ref{fig1}(a).

For the doped system, the case that a dopant replaces the atoms of undoped systems is discussed in our work. The dopant does not change the lattice constant and the potential energy of donor-doped and acceptor-doped semiconductors can be written as \cite{huang2017high}, 
\begin{equation}\label{Eq.1}
V(x)=\left\{\begin{array}{lll}
-V_0\left[1+\cos \left(2 \pi x / a_0\right)\right] & a \leqslant x \leqslant b \quad \text { or } \quad c \leqslant x \leqslant d \\
-(V_0+ \Delta V)\left[1+\cos \left(2 \pi x / a_0\right)\right] & b<x<c
\end{array} \quad\right.
\end{equation}
where $\Delta V=-0.13$ for acceptor-doped and $\Delta V=0.25$ for donor-doped. 
Because the dopant in the donor-doped system will contribute more electrons, the excess positive charge will deepen the potential energy. Thus $\Delta V$ is positive for the donor-doped system and negative for the dopant in the acceptor-doped system.

\begin{figure}[ht]
\includegraphics[width=8.50cm,angle=0]{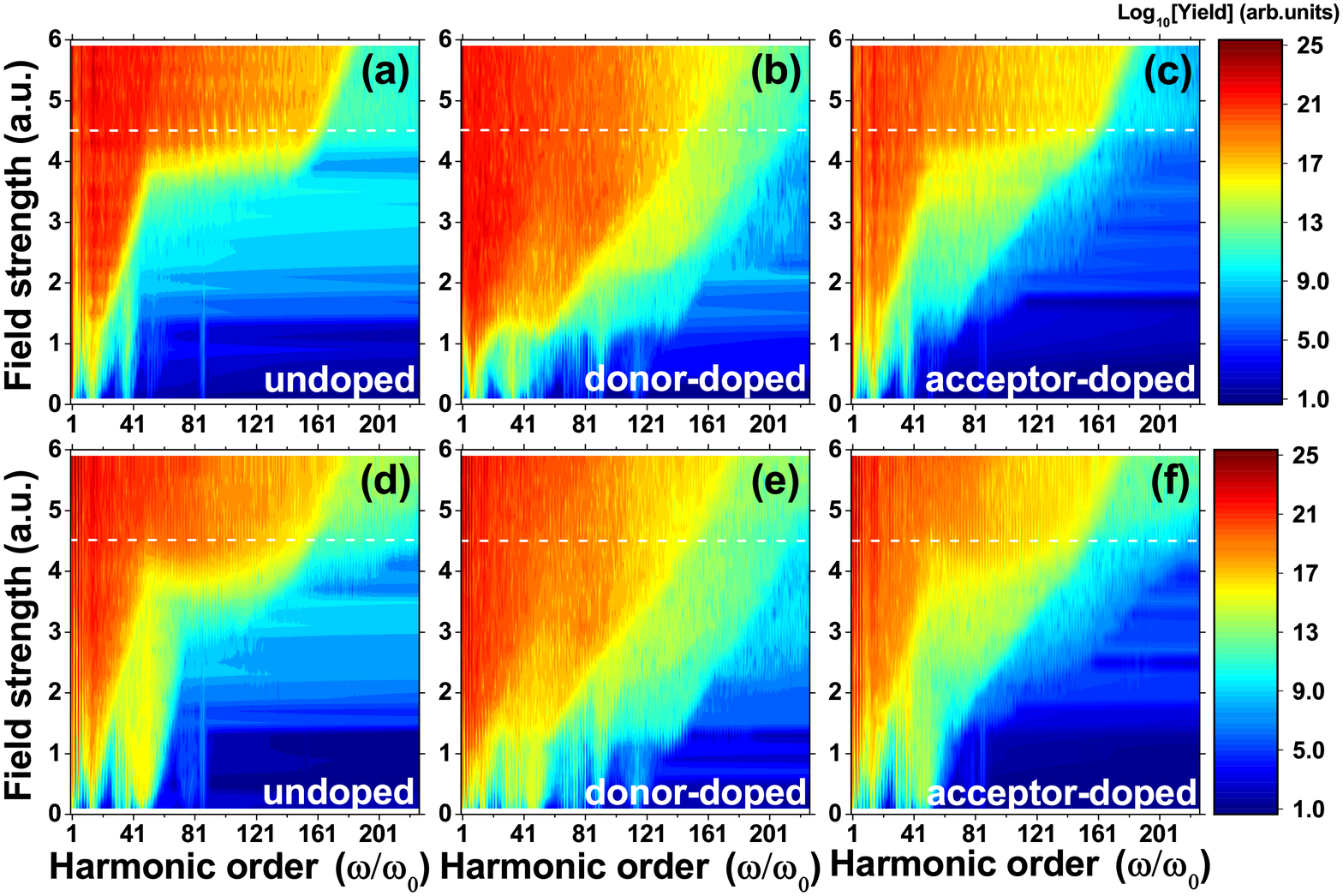}
\caption{\label{fig2}False-color representation of the harmonic spectra varies with the field strength $E_{0}$ (in logarithmic scale). The white dashed line corresponds to the $E_{0}$=0.0045 a.u. (a, d) undoped system, (b, e) donor-doped system, (c, f) acceptor-doped system. Upper row (a-c): the HHG spectra are obtained by Fourier transforming the current calculated by including only the highest occupied state. Bottom row (d-f): the HHG spectra are obtained by Fourier transforming the current calculated by including all occupied states.}
\end{figure}

The atomic doping density is defined as the ratio of the number of impurities to the total number of atoms. The typical atomic doping density of the bulk crystals in the experiment is between 0.1\% and 3\%. The band structure is not changed much at a low doping rate of around ~1\% \cite{yu2019enhanced}. In our work, the atomic doping density of 0.83\% is used from Fig. \ref{fig1} to Fig. \ref{fig6}. The HHG spectra for the doped system are calculated for one dopant atom in a finite chain with $N$=121 ions. For $N$=121 ions, the generated HHG spectra have all well-resolved structures compared with the HHG spectra from bulk system \cite{guan2016high,hansen2017finite-system,hansen2017high-order}. To test the convergence of results, the field-strength dependent calculation for HHG spectra has been also carried out for $N$=488 and $N$=1220 for comparison and the results are presented in the supplementary material.

In Fig. \ref{fig7}, HHG spectra in the doped system with different atomic doping densities of 1.64\%, 2.44\%, and 9.09\% are compared. For doping densities of 1.64\%, 2.44\%, and 9.09\%, the finite chain is constructed for 2 dopant atoms in a finite chain with $N$=122 ions, 4 dopant atoms in a finite chain with $N$=124 ions, 11 dopant atoms in a finite chain with $N$=121 ions, respectively. 

Figure \ref{fig1}(a) shows the potential energy of acceptor-doped (red dashed-dotted line), undoped (blue solid line), and donor-doped semiconductor (green dashed line), respectively. In Fig. \ref{fig1}(a), the eigenstate energy of the above atomic chain is obtained by solving the eigenstate wavefunction on a coordinate grid \cite{huang2017high}. The band structure in $k$-space can be obtained by the method of Bloch states expansion \cite{wu2015high,guan2016high,Wang2021Model} as shown in Fig. \ref{fig1}(b). After the eigenstate is obtained, the time-dependent calculation with the external driving laser field is solved in coordinate space by using the second-order split-operator method \cite{feit1982solution}. After the time-dependent wave function $\psi_i(t)$ is obtained, the laser-induced current contributed by the $i$th occupied state can be obtained by evaluating the expectation value of the momentum operator, $j_{i}(t)=\left\langle\psi_i(t)|\hat{p}| \psi_i(t)\right\rangle$. The current includes the contribution of all electrons can be expressed by,  
\begin{equation}\label{Eq.2}
j(t)=-\sum_{i=1}^N\left\langle\psi_i(t)|\hat{p}| \psi_i(t)\right\rangle,
\end{equation}
where $i$ is the eigenstate number and $N$ is the number of atoms. The corresponding HHG spectra contributed by $i$th occupied state or all occupied state can be obtained by performing the Fourier transform of the corresponding current, respectively.

The driving laser field used in our work has a Gaussian profile in the time domain and is given by,
$E(t)=E_0 \exp \left[-4 \ln (2) t^2 / \tau^2\right] \cos (\omega_{0} t),$ where $E_{0}$ is the amplitude of the driving field, $\omega_{0}$ is the fundamental frequency, and $\tau$ is the full width of half maximum (FWHM) of the laser field. The vector potential of the driving laser field is defined as $A(t)=-\int_{-\infty}^t E\left(t^{\prime}\right) d t^{\prime}$. Then the amplitude of the vector potential is given by, $A_{peak}=E_{0}/\omega_{0}$. And the wave vector of the electron is defined as \cite{du2017quasi-classical,jia2017nonadiabatic},
\begin{equation}\label{Eq.3}
k(t)=k_{0}+\frac{e}{\hbar} A(t),
\end{equation}  
where $k_{0}$ is the initial wave vector at $\Gamma$ point. When the electron-hole pairs are created through tunnel excitation, the excited carrier does intraband motion driven by the laser field. In the classical model, when electron and hole recombine and harmonic photons are emitted with the order given by, $\eta(t)=\frac{\varepsilon_{c}(k(t))-\varepsilon_{v}(k(t))}{\hbar \omega_{0}},$ where $\varepsilon_{c}(k(t))$ and $\varepsilon_{v}(k(t))$ are the energy of electrons in CB and holes in the VB, respectively.

\section{Results and discussion}
Figure \ref{fig1}(b) presents the energy band of the undoped system in the $k$-space. In Fig. \ref{fig1}(b), the blue, red, orange, purple, and cyan solid line shows the valence band (VB), conduction band 1 (CB1), conduction band 2 (CB2), conduction band 3 (CB3) and conduction band 4 (CB4), respectively. 

Figures \ref{fig1}(c) and \ref{fig1}(d) illustrate the HHG spectra driven by an eight-cycle Gaussian pulse with field strength $E_{0}$=0.0038 a.u. and central wavelength of 4000 nm. The vertical blue, red, orange, and purple dashed lines in Figs. \ref{fig1}(c-f) denote the minimum bandgap between CB1 and VB, the bandgap energy between CB1 and VB at the boundary of BZ which lead to the cutoff energy of the 1st plateau, the cutoff energy of the 2nd plateau, and the cutoff energy of the 3rd plateau, respectively.

In Figs. \ref{fig1}(c) and \ref{fig1}(d), the same as Huang $et. al$'s work, the second plateau of HHG spectra from donor-doped semiconductor (green solid line) is enhanced by several orders than its counterpart from the acceptor-doped and undoped system. The results by including only the highest occupied state and all electrons show the same trend. While the HHG spectra from acceptor-doped semiconductors are slightly stronger than undoped semiconductors. 
The large enhancement of the HHG spectra from donor-doped accords with Yu $et. al$'s work by using the TDDFT calculation \cite{yu2019enhanced} and the experimental observation of enhancement of HHG spectra in Cr-doped MgO \cite{nefedova2021enhanced}.  

In Figs. \ref{fig1}(e) and \ref{fig1}(f), the HHG spectra are driven by the same Gaussian laser field as the middle row except for a larger field strength $E_{0}$=0.0045 a.u. This field strength corresponds to $A_{peak}$=$\pi$/$a_{0}$ which is equal to the half-width of the Brillouin zone. In this case, the magnitude of the vector potential is able to drive the electron to reach the boundary of the BZ. Different from the case of $E_{0}$=0.0038 a.u., the yield of HHG spectra in the undoped system is larger than in the acceptor-doped and donor-doped systems.

Figure \ref{fig2} illustrates the HHG spectra vary with the field strength of the driving field. The HHG spectra shown in the upper row of Fig. \ref{fig2}(a-c) are calculated by only including the highest occupied state. While the bottom row presents the results calculated by including all the occupied states. The white dashed line marks the driving field strength $E_{0}$=0.0045 a.u. which corresponds to $A_{peak}=\pi/a_{0}$.

In Figs. \ref{fig2}(a) and \ref{fig2}(d), when $E_{0}$=0.0045 a.u., the HHG spectra from undoped semiconductor shows a double-plateau structure \cite{trajectory2017ikemachi}. For $A_{peak}=\pi/a_{0}$, the wave vector of electron $k(t)=k_{0}+A(t)$ can reach the boundary of Brillouin zone (BZ). The electrons in CB1 can be pumped to the CB2 through the band climbing mechanism \cite{trajectory2017ikemachi} and then the electrons do intraband motion driven by the laser field. In the following half optical cycle, the electrons are driven backward and move towards the center of BZ ($k$=0) and can be pumped to CB3 through band climbing. Therefore, the 2nd plateau and higher plateau structure appear when $A_{peak}=a_{0}/\omega_{0}$.

However, in Figs. \ref{fig2}(b) and \ref{fig2}(e), for donor-doped system, when the $A_{peak}$ is still below $a_{0}/\omega_{0}$, the 2nd plateau and higher order emission are clearly observed. The doped impurity energy level between the VB and CB1 is not occupied by electrons. Compared with the occupied states in the VB, the impurity energy level is closer to the CB1 and is easier to be excited to the CB1. The ladder provided by the impurity energy level increases the excitation rates of optical transition and strengthens the HHG dynamics \cite{yu2019enhanced,zhao2021impact}.

Figures \ref{fig2}(c) and \ref{fig2}(f) present the field-strength dependence of HHG spectra in acceptor-doped semiconductors. The same as the donor-doped system in the middle column, when the $A_{peak}$ is still below $a_{0}/\omega_{0}$, the 2nd plateau and higher order emission are observed. Although, in acceptor-doped semiconductors, the impurity energy level between the VB and CB1 is occupied by the electrons. However, the impurity energy level is close to the VB. Compared with the undoped semiconductor, the electrons in the VB are easier to excite to the CB1 under the same electric field strength. This makes the yield of HHG spectra stronger than the undoped system when the amplitude of the vector potential is below $a_{0}/\omega_{0}$. However, because the doped energy level in the acceptor-doped system is not occupied by the electrons, the acceptor-doped semiconductor does not have a channel of HHG contributed by the excitation from the impurity energy level to CB1 directly. This causes the yield of HHG from the acceptor-doped system to be less efficient than the donor-doped. 

\begin{figure}[ht]
\includegraphics[width=8.50cm,angle=0]{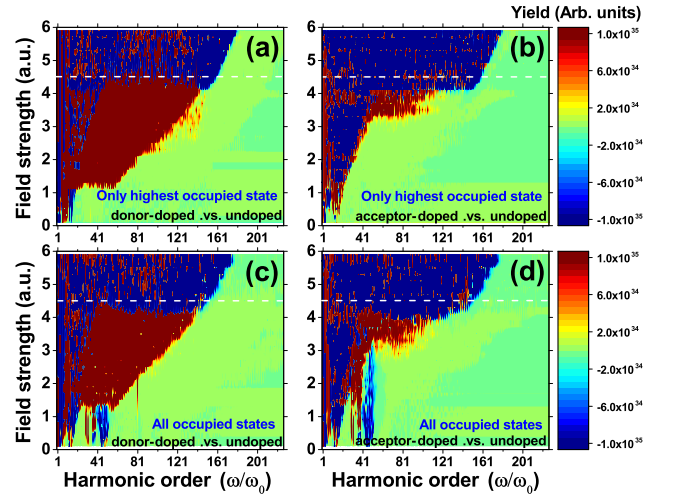}
\caption{\label{fig3}Left column: the HHG yield difference obtained by subtracting yield from undoped to the donor-doped system. Right column: the HHG yield difference obtained by subtracting yield from undoped to the acceptor-doped system. The red-most indicates HHG yield from the doped system is stronger than the undoped. While the blue-most shows the HHG yield from the undoped system is larger. (a) The difference between Figs. \ref{fig2}(b) and \ref{fig2}(a). (b) The difference between Figs. \ref{fig2}(c) and \ref{fig2}(a). (c) The difference between Figs. \ref{fig2}(e) and \ref{fig2}(d). (d) The difference between Figs. \ref{fig2}(f) and \ref{fig2}(d). The yield difference is plotted on a linear scale.}
\end{figure}

In Fig. \ref{fig3}, the effect of doping on HHG is investigated by subtracting the yield of HHG spectra in the undoped system from the doped system. Figures \ref{fig3}(a) and \ref{fig3}(c) show the difference between the HHG spectra from the donor-doped and the undoped system. Figures \ref{fig3}(b) and \ref{fig3}(d) present the difference between the HHG spectra from the acceptor-doped and the undoped system. The red-most presents that the yield in the doped system is higher than the undoped, while the blue-most presents that the yield in the undoped system is higher than the doped system. Both the calculation by including only the highest occupied state and including all occupied states shows the division separated by the white-dashed line corresponds to the field strength of $E_{0}$=0.0045 a.u. ($A_{peak}=a_{0}/\omega_{0}$). When $A_{peak}$ is below $a_{0}/\omega_{0}$, the HHG from doped system is stronger. In contrast, for $A_{peak}$ is around or above $a_{0}/\omega_{0}$, the HHG from the undoped system is brighter. The result indicates the enhancement of yield in HHG by doping is field strength dependent for typical atomic doping density around 1\% in the experiment.

\begin{figure*}[ht]
\includegraphics[width=16.0cm,angle=0]{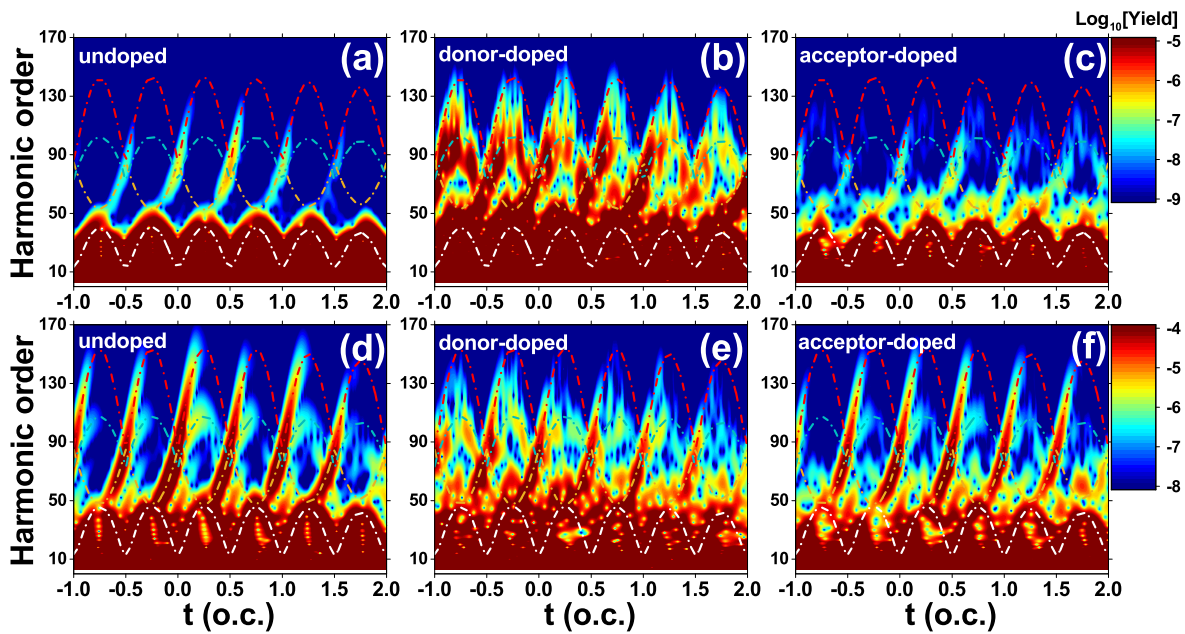}
\caption{\label{fig4}The temporal profile of HHG spectra plotted in logarithmic scale. Upper row (a-c): the temporal profile of HHG driven by laser field with $E_{0}$=0.0038 a.u. Bottom row (d-f): the temporal profile of HHG driven by laser field with $E_{0}$=0.0045 a.u. (a, d) undoped system, (b, e) donor-doped system, (c, f) acceptor-doped system. The white, orange, and red dashed-dotted lines show the trajectory predicted by the quasi-classical model for the interband transition from the conduction band CB1, CB2, and CB3 to VB, respectively. The cyan dashed-dotted lines represent the trajectory involving the transition between CB3 and CB1. When the driving field strength $E_{0}$=0.0038 a.u., the harmonic yield at the second and third plateau of the donor-doped semiconductor becomes much stronger than the undoped target and acceptor-doped system. For higher field strength $E_{0}$=0.0045 a.u., for the doped target, the harmonic in the second and third plateau become weaker than the undoped system.}
\end{figure*}


In Fig. \ref{fig4}, the temporal profile of HHG is calculated by including all occupied states. Figures \ref{fig4}(a), \ref{fig4}(b), and \ref{fig4}(c) in the upper row present the HHG driven by an eight-cycle, 4000 nm laser with a field strength $E_{0}$=0.0038 a.u. from undoped, donor-doped, and acceptor-doped systems, respectively. In Fig. \ref{fig4}(b), the trajectory of donor-doped systems is stronger than that of undoped and acceptor-doped systems in Figs. \ref{fig4}(a) and \ref{fig4}(c), respectively.

In Figs. \ref{fig4}(d), \ref{fig4}(e) and \ref{fig4}(f), for driving field strength of $E_{0}$=0.0045 a.u., the temporal profile of HHG in undoped semiconductor in Fig. \ref{fig4}(d) is stronger than that of the donor-doped and acceptor-doped system in Figs. \ref{fig4}(e) and \ref{fig4}(f), respectively. In addition, the HHG trajectories in Fig. \ref{fig4}(d) are almost perfectly repeated in each half optical cycle and have a well-resolved temporal structure. This indicates these trajectories will interfere with each other constructively which leads to the enhancement of the total HHG spectra in the frequency domain.

\begin{figure}[ht!]
\centering\includegraphics[width=8.5cm]{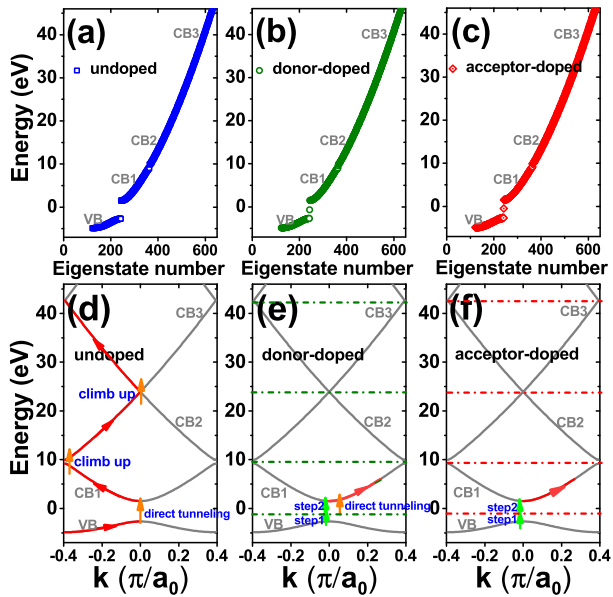}
\caption{\label{fig5}The energy band structure of the undoped (a, d), donor-doped (b, e), and acceptor-doped system (c, f) including a valence band VB1 and conduction bands CB1, CB2, CB3, and all impurity energy levels. Upper row(a-c): The band energies are plotted in ascending order. Bottom row(d-f): Pictorial representation of the band structure and electron trajectories in momentum-space.}
\end{figure}

In Fig. \ref{fig5}, the left, middle, and right columns show the band structure of undoped, donor-doped, and acceptor-doped systems, respectively. The upper row presents the eigenstate energy in ascending order including the energy band VB, CB1, CB2, CB3, and all doped impurity energy levels. The "in-band" energies in Figs. \ref{fig5}(b) and \ref{fig5}(c) for the donor-doped system are almost unchanged compared with the undoped system as shown in Fig. \ref{fig5}(a) \cite{yu2019enhanced}. In Fig. \ref{fig5}(b), for the donor-doped system, the impurity energy level between VB and CB1 is occupied by electrons. While for the acceptor-doped system in Fig. \ref{fig5}(c), the impurity energy level between VB and CB1 is unoccupied. 

Figures \ref{fig5}(d), \ref{fig5}(e), and \ref{fig5}(f) are the pictorial representation of the band structure and laser field-induced electron trajectories of undoped, donor-doped, and acceptor-doped systems in $k$-space, respectively. For the donor-doped system in Fig. \ref{fig5}(e), the electrons can be directly pumped from the occupied impurity energy level to the CB1, which is marked by the upward orange arrow between VB and CB1 in Fig. \ref{fig5}(e). According to the Keldysh model \cite{keldysh1965ionization}, the excitation rate increases exponentially with the narrowing of the band gap. The occupied doped energy level which is closer to CB1 has a much larger excitation rate to the CB1 compared with the other occupied states in the VB.  


When the field strength of the driving field is weak ($A_{peak}<a_{0}$/$\omega_{0}$), for the donor-doped system, the direct excitation from impurity energy level is the main factor which causes the enhancement of the HHG dynamics by several orders. For the acceptor-doped system in Fig. \ref{fig5}(e), the unoccupied impurity states between VB and CB1 provide a "ladder" in the step-by-step transition process which results in an increase of HHG. 

The orange arrows between CB1 and CB2 in Fig. \ref{fig5}(d) denote the "band-climbing" process \cite{trajectory2017ikemachi}. When the field strength of the driving laser is large ($A_{peak}> \pi/a_{0}$).  After the transition to CB1 from VB, the electrons do intraband motions driven by the laser and can be accelerated to the BZ boundary and can climb up to the higher CB2. In the following half optical cycle, the electrons pumped to CB2 are driven backward by the laser field and do intraband motion along CB2. When the electrons in CB2 are driven to the vicinity of $k$=0, the electrons can climb up to CB3. The "band-climbing" process has been proposed to explain the multi-plateau structure in solid HHG spectra \cite{trajectory2017ikemachi}. In our work, when $A_{peak}= \pi/a_0$, the electrons in the undoped semiconductor can be pumped to a CB2 and then CB3 through a "band-climbing" process, which leads to a large enhancement of yield in the second and third plateau on HHG spectra.

\begin{figure}[ht!]
\centering\includegraphics[width=8.5cm]{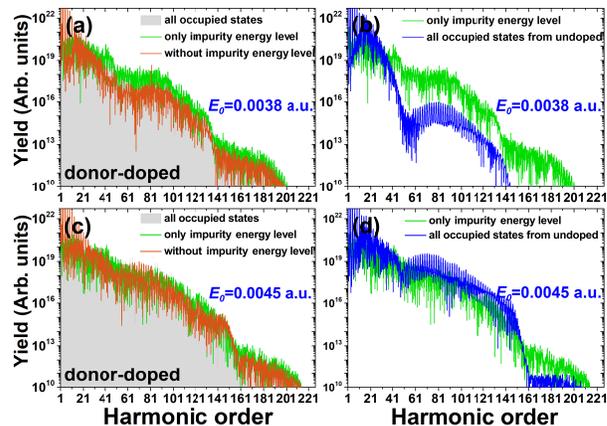}
\caption{\label{fig6}(a, c) The HHG spectra for the donor-doped system driven by laser with field strength $E_{0}$=0.0038 a.u. (a) and $E_{0}$=0.0045 a.u. (c), respectively. The gray, red, and green solid line represents HHG obtained by Fourier transforming the total current from all occupied states, with only the highest doped energy level state, and without this state, respectively. (b, d) The comparison of the HHG spectra contributed by the doped energy level, with the field strength of driving field $E_{0}$=0.0038 a.u. in (b) and $E_{0}$=0.0045 a.u. in (d), respectively.}
\end{figure}

In Fig. \ref{fig6}, we analyze the HHG spectra contributed by the doped impurity energy level. The left column of Fig. \ref{fig6} presents the HHG spectra from the donor-doped system. The right column of Fig. \ref{fig6} compares the HHG spectra from the impurity energy level in the donor-doped system and HHG spectra from the undoped system by including all occupied states.

In Figs. \ref{fig6}(a) and \ref{fig6}(c), the gray solid line represents the HHG spectra contributed by all occupied states. The green solid line and orange solid line represent the HHG spectra contributed by only the impurity energy level and the other occupied states, respectively. In Fig. \ref{fig6}(a), for driving field strength $E_{0}$=0.0038 a.u., in a wide range of orders, the HHG spectra obtained from the highest occupied state agrees with the spectra including all occupied states. This accords with Yu $et. al$'s work \cite{yu2019enhanced} that HHG spectra from the donor-doped system are mainly contributed by the impurity energy level. 

In Fig. \ref{fig6}(b), HHG spectra contributed by only the impurity energy level are larger than the HHG spectra from the undoped system by including all electrons. This indicates enhancement of the yield of HHG by donor-doping is mainly caused by the impurity energy level. 

In Fig. \ref{fig6}(c), driven by the laser with field strength $E_{0}$=0.0045 a.u., the yield of HHG spectra contributed by all the other states is comparable with the HHG contributed by only the impurity energy level. As discussed above, when the amplitude of vector potential reaches $a_{0}/\omega_{0}$, a large number of electrons can be driven to the vicinity of the boundary of BZ and pumped to a CB2 through the "band climbing".

In Fig. \ref{fig6}(d), the HHG spectra from the undoped system are stronger than the HHG spectra contributed by the impurity energy level from the donor-doped system. In addition, the HHG spectra from the undoped system have a clear harmonic signal of integer order. This indicates the HHG trajectories in the time domain have constructive interference with each other. 

\begin{figure}[ht!]
\centering\includegraphics[width=8.5cm]{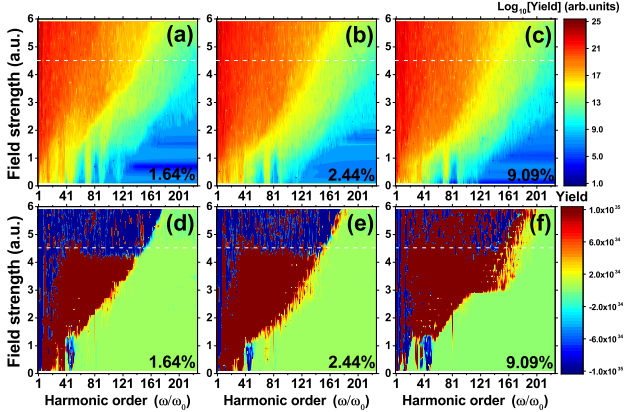}
\caption{\label{fig7}(a-c) False-color representation of the harmonic spectra varies with the field strength $E_{0}$ for the doping rate of 1.64\% (a), 2.44\% (b) and 9.09\% (c), respectively. The HHG spectra are obtained by Fourier transforming the current calculated with all occupied states. The white dashed line corresponds to the $E_{0}$=0.0045 a.u. (d-f) The difference between HHG yield from the donor-doped and undoped system. The laser parameters are the same as those used in Fig. \ref{fig4}.}
\end{figure}

Further, the effect of atomic doping density on HHG spectra is investigated. Figures \ref{fig7}(a), \ref{fig7}(b) and \ref{fig7}(c) represent the HHG from the donor-doped semiconductor with the atomic doping density of 1.64\%, 2.44\% and 9.09\%, respectively. Figures \ref{fig7}(d), \ref{fig7}(e) and \ref{fig7}(f) represent the difference between the HHG spectra from the donor-doped system and the undoped system. With the increase of the atomic doping density, the channels transited directly from the occupied impurity energy level to the CB1 will increase, which will give rise to the enhancement of the HHG.

In Fig. \ref{fig7}(d), when the atomic doping density is equal to 1.64\%, the enhancement of HHG by doping shows clearly dependence on the laser field strength. An atomic doping density of 1.64\% should be common in experiments. When the atomic doping density increases to 2.44\% and 9.09\%, with the increase of impurity energy levels, for $A_{peak}$ above the $\pi/a_{0}$, the advantage of an undoped system become less obvious. However, such a high doping density of 9.09\% is rare in usual experiments. 

\section{Conclusion}
In conclusion, our research shows that under the common atomic doping density of 0.1$\%$-3$\%$, the improvement of harmonic yield in the doped system such as nanomaterials or bulk materials is field strength dependent. When the amplitude of the driving laser vector potential $A_{peak}$ is lower than $\pi$/$a_{0}$, because the impurity energy level provides a 'ladder' for interband transition, it makes electrons easier to be transited from VB to CB1, CB2, and CB3, which cause the harmonic yield of the acceptor-doped and donor-doped system to be larger than that of the undoped system. In particular, in a donor-doped system, the electrons from impurity energy levels can be directly excited to the CB1, which greatly improves the harmonic yield.

When $A_{peak}$ is around $\pi/a_{0}$ or higher than $\pi/a_{0}$, because the electrons in CB1 can move to the boundary of BZ, the electrons can climb up to the CB2, and then do intraband motions in the CB2 driven by the external field, and then be pumped to the higher CB3 through the band climbing near $k$=0, which greatly enhances the interband transition \cite{trajectory2017ikemachi}. Moreover, since there is no impurity energy level in the undoped system, a single channel of HHG makes the trajectories from the undoped system have a period of half optical cycle in the time domain. They interfere with each other constructively which makes the harmonic yield in the frequency domain to be enhanced. With the increase of atomic doping density, the yield of HHG from the doped system can be increased further which will reduce this field strength dependence.

\begin{acknowledgments}
We thanks the support from the National Natural Science Foundation of China (No. 12104395), and Zhejiang Provincial Natural Science Foundation of China (No. LQ22A040004). 
\end{acknowledgments}

\appendix

\section{Convergence Tests}
\begin{figure}[ht!]
\centering\includegraphics[width=8.5cm]{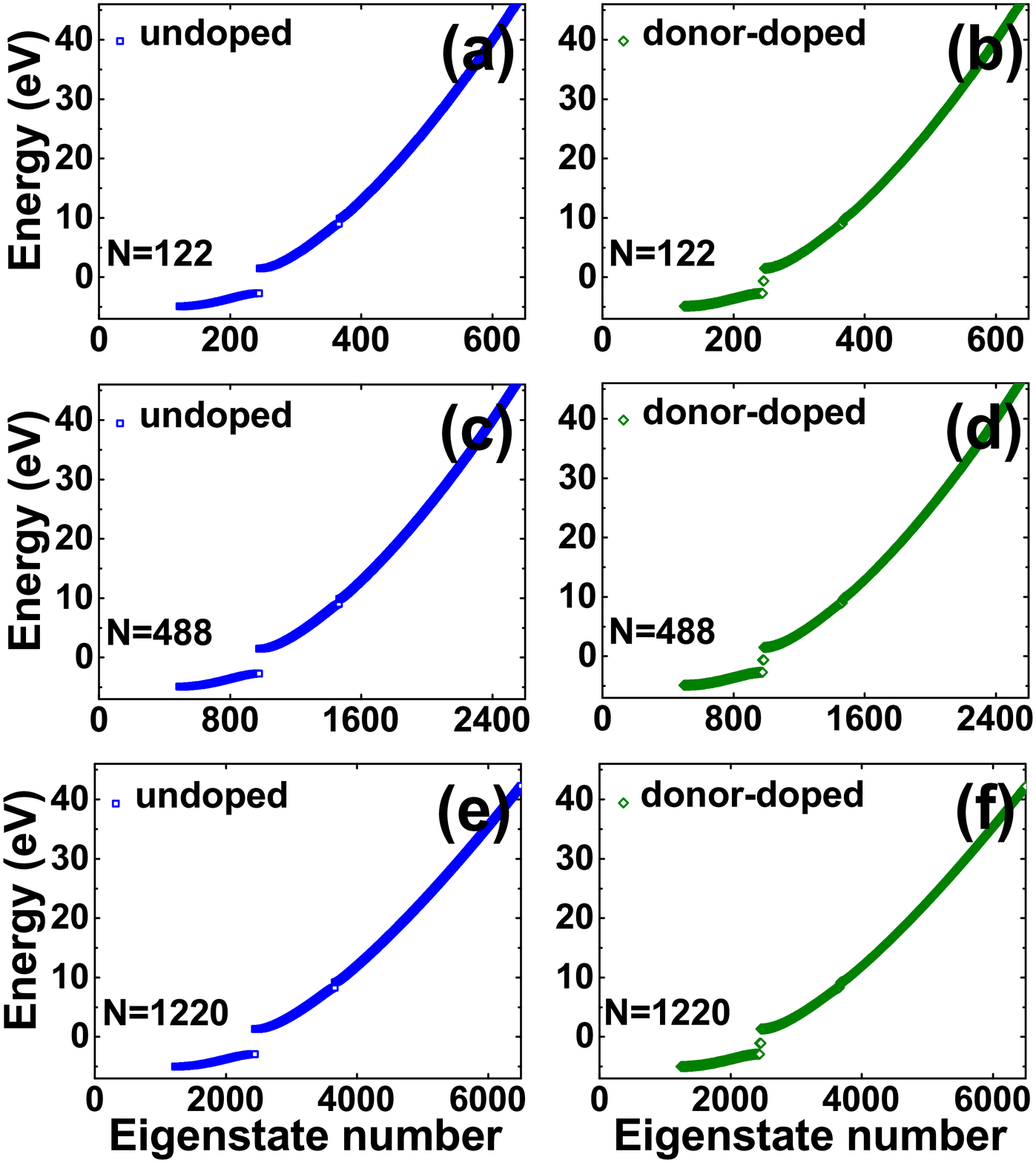}
\caption{\label{fig8}The energy band structure of the undoped (a, c, e) and donor-doped (b, d, f) including a valence band VB1 and conduction bands CB1, CB2, CB3, and all impurity energy levels. The band energies are plotted in ascending order. The system sizes of $N$=122, 488, and 1220 are presented with growing system sizes from top to bottom.}
\end{figure}

\begin{figure}[ht!]
\centering\includegraphics[width=8.5cm]{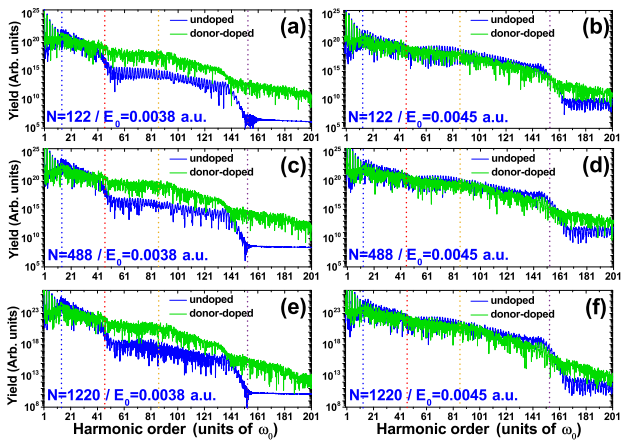}
\caption{\label{fig9}HHG spectra for system sizes of $N$=122, 488, and 1220 using the same laser parameters as in Fig. \ref{fig1} in the manuscript. The HHG spectra are presented with a growing system size from top to bottom to test the convergence of our calculation results. The vertical blue, red, orange and purple dashed lines present the minimum bandgap between CB1 and VB, the bandgap energy between CB1 and VB at the boundary of BZ which lead to the cutoff energy of the 1st plateau, the cutoff energy of the 2nd plateau, and the cutoff energy of the 3rd plateau, respectively. In the left column (a, c, e), when the amplitude of the vector potential $A_{peak}$ is below the $\pi$/$a_{0}$ ($E_0$=0.0038 a.u.), the yield of HHG from the donor-doped system is several orders larger than the undoped system. In the right column (b, d, f), for $A_{peak}$ reaches the half-width of the BZ ($E_0$=0.0045 a.u.), on the contrary, the yield of HHG from the undoped system is larger than the donor-doped system. Since the excitation for crystal momentum states far from the minimal band gap is low, the total HHG spectra in this figure include the crystal momentum states that are located within 5$\%$ distance from the minimum band gap.}
\end{figure}

Figures \ref{fig8} and \ref{fig9} present the convergence test of our calculation results. Figure \ref{fig8} shows the energy band structure of the undoped (a, c, e) and donor-doped systems (b, d, f) of a finite chain in ascending order, respectively. For doping densities of 1.64$\%$, the finite chain with system sizes of $N$=122, $N$=488, and $N$=1220 is constructed for 2 dopant atoms in a finite chain with $N$=122 ions, 8 dopant atoms in a finite chain with $N$=488 ions, 20 dopant atoms in a finite chain with $N$=1220 ions, respectively. 

For instance, in Fig. \ref{fig8}(a), for the undoped system with system sizes of $N$=122, the VB, CB1, CB2, CB3 corresponds to the state numbers 123-244, 245-366, 367-488, 489-610, respectively. In Fig. \ref{fig8}(b), for the donor-doped system with system sizes of $N$=122, the VB, CB1, CB2, and CB3 correspond to the state numbers 125-244, 247-366, 369-488, and 491-610, respectively. The state numbers 245 and 246 correspond to the occupied impurity energy states between VB and CB1. Figs. \ref{fig8}(c) and \ref{fig8}(d) presents the band energies of undoped and donor-doped systems with system sizes of $N$=488, respectively. Figs. \ref{fig8}(e) and \ref{fig8}(f) are system sizes of $N$=1220. 

Figure \ref{fig9} shows the comparison of HHG spectra from undoped and donor-doped systems for sizes of $N$=122 (a, b), 488 (c, d), and 1220 (e, f) using the same laser parameters as Fig. \ref{fig1}. The HHG spectra are presented with growing system size from top to bottom to test the convergence of our calculation results. Since the excitation for crystal momentum states far from the minimal band gap is low, the total HHG spectra in Fig. \ref{fig9} include the crystal momentum states that are located within 5$\%$ distance from the minimum band gap. For system sizes of $N$=122 (a, b), $N$=488 (c, d), and $N$=1220 (e, f), we observe nearly identical HHG spectra. In the left column of Figs. \ref{fig9}(a, c, e), when the amplitude of the vector potential $A_{peak}$ is below the $\pi$/a0 ($E_0$=0.0038 a.u.), the yield of HHG from the donor-doped system is several orders larger than the undoped system. In the right column (b, d, f), for $A_{peak}$ reaches the half-width of the BZ ($E_0$=0.0045 a.u.), on the contrary, the yield of HHG from the undoped system is larger than the donor-doped one. Our results show, under the conventional doping ratio of 0.5$\%$-3$\%$, the enhancement of HHG by doping is field strength dependent in doped systems such as nano-sized or bulk materials.

\bibliography{apssamp}

\begin{thebibliography}{47}%
\makeatletter
\providecommand \@ifxundefined [1]{%
 \@ifx{#1\undefined}
}%
\providecommand \@ifnum [1]{%
 \ifnum #1\expandafter \@firstoftwo
 \else \expandafter \@secondoftwo
 \fi
}%
\providecommand \@ifx [1]{%
 \ifx #1\expandafter \@firstoftwo
 \else \expandafter \@secondoftwo
 \fi
}%
\providecommand \natexlab [1]{#1}%
\providecommand \enquote  [1]{``#1''}%
\providecommand \bibnamefont  [1]{#1}%
\providecommand \bibfnamefont [1]{#1}%
\providecommand \citenamefont [1]{#1}%
\providecommand \href@noop [0]{\@secondoftwo}%
\providecommand \href [0]{\begingroup \@sanitize@url \@href}%
\providecommand \@href[1]{\@@startlink{#1}\@@href}%
\providecommand \@@href[1]{\endgroup#1\@@endlink}%
\providecommand \@sanitize@url [0]{\catcode `\\12\catcode `\$12\catcode
  `\&12\catcode `\#12\catcode `\^12\catcode `\_12\catcode `\%12\relax}%
\providecommand \@@startlink[1]{}%
\providecommand \@@endlink[0]{}%
\providecommand \url  [0]{\begingroup\@sanitize@url \@url }%
\providecommand \@url [1]{\endgroup\@href {#1}{\urlprefix }}%
\providecommand \urlprefix  [0]{URL }%
\providecommand \Eprint [0]{\href }%
\providecommand \doibase [0]{https://doi.org/}%
\providecommand \selectlanguage [0]{\@gobble}%
\providecommand \bibinfo  [0]{\@secondoftwo}%
\providecommand \bibfield  [0]{\@secondoftwo}%
\providecommand \translation [1]{[#1]}%
\providecommand \BibitemOpen [0]{}%
\providecommand \bibitemStop [0]{}%
\providecommand \bibitemNoStop [0]{.\EOS\space}%
\providecommand \EOS [0]{\spacefactor3000\relax}%
\providecommand \BibitemShut  [1]{\csname bibitem#1\endcsname}%
\let\auto@bib@innerbib\@empty
\bibitem [{\citenamefont {McPherson}\ \emph {et~al.}(1987)\citenamefont
  {McPherson}, \citenamefont {Gibson}, \citenamefont {Jara}, \citenamefont
  {Johann}, \citenamefont {Luk}, \citenamefont {McIntyre}, \citenamefont
  {Boyer},\ and\ \citenamefont {Rhodes}}]{mcpherson1987studies}%
  \BibitemOpen
  \bibfield  {author} {\bibinfo {author} {\bibfnamefont {A.}~\bibnamefont
  {McPherson}}, \bibinfo {author} {\bibfnamefont {G.}~\bibnamefont {Gibson}},
  \bibinfo {author} {\bibfnamefont {H.}~\bibnamefont {Jara}}, \bibinfo {author}
  {\bibfnamefont {U.}~\bibnamefont {Johann}}, \bibinfo {author} {\bibfnamefont
  {T.~S.}\ \bibnamefont {Luk}}, \bibinfo {author} {\bibfnamefont
  {I.}~\bibnamefont {McIntyre}}, \bibinfo {author} {\bibfnamefont
  {K.}~\bibnamefont {Boyer}},\ and\ \bibinfo {author} {\bibfnamefont {C.~K.}\
  \bibnamefont {Rhodes}},\ }\bibfield  {title} {\bibinfo {title} {Studies of
  multiphoton production of vacuum-ultraviolet radiation in the rare gases},\
  }\href@noop {} {\bibfield  {journal} {\bibinfo  {journal} {JOSA B}\ }\textbf
  {\bibinfo {volume} {4}},\ \bibinfo {pages} {595} (\bibinfo {year}
  {1987})}\BibitemShut {NoStop}%
\bibitem [{\citenamefont {Corkum}(1993)}]{corkum1993plasma}%
  \BibitemOpen
  \bibfield  {author} {\bibinfo {author} {\bibfnamefont {P.~B.}\ \bibnamefont
  {Corkum}},\ }\bibfield  {title} {\bibinfo {title} {Plasma perspective on
  strong field multiphoton ionization},\ }\href@noop {} {\bibfield  {journal}
  {\bibinfo  {journal} {Physical Review Letters}\ }\textbf {\bibinfo {volume}
  {71}},\ \bibinfo {pages} {1994} (\bibinfo {year} {1993})}\BibitemShut
  {NoStop}%
\bibitem [{\citenamefont {Lewenstein}\ \emph {et~al.}(1994)\citenamefont
  {Lewenstein}, \citenamefont {Balcou}, \citenamefont {Ivanov}, \citenamefont
  {L’huillier},\ and\ \citenamefont {Corkum}}]{lewenstein1994theory}%
  \BibitemOpen
  \bibfield  {author} {\bibinfo {author} {\bibfnamefont {M.}~\bibnamefont
  {Lewenstein}}, \bibinfo {author} {\bibfnamefont {P.}~\bibnamefont {Balcou}},
  \bibinfo {author} {\bibfnamefont {M.~Y.}\ \bibnamefont {Ivanov}}, \bibinfo
  {author} {\bibfnamefont {A.}~\bibnamefont {L’huillier}},\ and\ \bibinfo
  {author} {\bibfnamefont {P.~B.}\ \bibnamefont {Corkum}},\ }\bibfield  {title}
  {\bibinfo {title} {Theory of high-harmonic generation by low-frequency laser
  fields},\ }\href@noop {} {\bibfield  {journal} {\bibinfo  {journal} {Physical
  Review A}\ }\textbf {\bibinfo {volume} {49}},\ \bibinfo {pages} {2117}
  (\bibinfo {year} {1994})}\BibitemShut {NoStop}%
\bibitem [{\citenamefont {Paul}\ \emph {et~al.}(2001)\citenamefont {Paul},
  \citenamefont {Toma}, \citenamefont {Breger}, \citenamefont {Mullot},
  \citenamefont {Aug{\'e}}, \citenamefont {Balcou}, \citenamefont {Muller},\
  and\ \citenamefont {Agostini}}]{paul2001observation}%
  \BibitemOpen
  \bibfield  {author} {\bibinfo {author} {\bibfnamefont {P.-M.}\ \bibnamefont
  {Paul}}, \bibinfo {author} {\bibfnamefont {E.~S.}\ \bibnamefont {Toma}},
  \bibinfo {author} {\bibfnamefont {P.}~\bibnamefont {Breger}}, \bibinfo
  {author} {\bibfnamefont {G.}~\bibnamefont {Mullot}}, \bibinfo {author}
  {\bibfnamefont {F.}~\bibnamefont {Aug{\'e}}}, \bibinfo {author}
  {\bibfnamefont {P.}~\bibnamefont {Balcou}}, \bibinfo {author} {\bibfnamefont
  {H.~G.}\ \bibnamefont {Muller}},\ and\ \bibinfo {author} {\bibfnamefont
  {P.}~\bibnamefont {Agostini}},\ }\bibfield  {title} {\bibinfo {title}
  {Observation of a train of attosecond pulses from high harmonic generation},\
  }\href@noop {} {\bibfield  {journal} {\bibinfo  {journal} {Science}\ }\textbf
  {\bibinfo {volume} {292}},\ \bibinfo {pages} {1689} (\bibinfo {year}
  {2001})}\BibitemShut {NoStop}%
\bibitem [{\citenamefont {Hentschel}\ \emph {et~al.}(2001)\citenamefont
  {Hentschel}, \citenamefont {Kienberger}, \citenamefont {Spielmann},
  \citenamefont {Reider}, \citenamefont {Milosevic}, \citenamefont {Brabec},
  \citenamefont {Corkum}, \citenamefont {Heinzmann}, \citenamefont {Drescher},\
  and\ \citenamefont {Krausz}}]{hentschel2001attosecond}%
  \BibitemOpen
  \bibfield  {author} {\bibinfo {author} {\bibfnamefont {M.}~\bibnamefont
  {Hentschel}}, \bibinfo {author} {\bibfnamefont {R.}~\bibnamefont
  {Kienberger}}, \bibinfo {author} {\bibfnamefont {C.}~\bibnamefont
  {Spielmann}}, \bibinfo {author} {\bibfnamefont {G.~A.}\ \bibnamefont
  {Reider}}, \bibinfo {author} {\bibfnamefont {N.}~\bibnamefont {Milosevic}},
  \bibinfo {author} {\bibfnamefont {T.}~\bibnamefont {Brabec}}, \bibinfo
  {author} {\bibfnamefont {P.}~\bibnamefont {Corkum}}, \bibinfo {author}
  {\bibfnamefont {U.}~\bibnamefont {Heinzmann}}, \bibinfo {author}
  {\bibfnamefont {M.}~\bibnamefont {Drescher}},\ and\ \bibinfo {author}
  {\bibfnamefont {F.}~\bibnamefont {Krausz}},\ }\bibfield  {title} {\bibinfo
  {title} {Attosecond metrology},\ }\href@noop {} {\bibfield  {journal}
  {\bibinfo  {journal} {Nature}\ }\textbf {\bibinfo {volume} {414}},\ \bibinfo
  {pages} {509} (\bibinfo {year} {2001})}\BibitemShut {NoStop}%
\bibitem [{\citenamefont {Lein}\ \emph {et~al.}(2002)\citenamefont {Lein},
  \citenamefont {Marangos},\ and\ \citenamefont {Knight}}]{lein2002electron}%
  \BibitemOpen
  \bibfield  {author} {\bibinfo {author} {\bibfnamefont {M.}~\bibnamefont
  {Lein}}, \bibinfo {author} {\bibfnamefont {J.~P.}\ \bibnamefont {Marangos}},\
  and\ \bibinfo {author} {\bibfnamefont {P.~L.}\ \bibnamefont {Knight}},\
  }\bibfield  {title} {\bibinfo {title} {Electron diffraction in
  above-threshold ionization of molecules},\ }\href
  {https://doi.org/10.1103/PhysRevA.66.051404} {\bibfield  {journal} {\bibinfo
  {journal} {Phys. Rev. A}\ }\textbf {\bibinfo {volume} {66}},\ \bibinfo
  {pages} {051404} (\bibinfo {year} {2002})}\BibitemShut {NoStop}%
\bibitem [{\citenamefont {Itatani}\ \emph {et~al.}(2004)\citenamefont
  {Itatani}, \citenamefont {Levesque}, \citenamefont {Zeidler}, \citenamefont
  {Niikura}, \citenamefont {P{\'e}pin}, \citenamefont {Kieffer}, \citenamefont
  {Corkum},\ and\ \citenamefont {Villeneuve}}]{itatani2004tomographic}%
  \BibitemOpen
  \bibfield  {author} {\bibinfo {author} {\bibfnamefont {J.}~\bibnamefont
  {Itatani}}, \bibinfo {author} {\bibfnamefont {J.}~\bibnamefont {Levesque}},
  \bibinfo {author} {\bibfnamefont {D.}~\bibnamefont {Zeidler}}, \bibinfo
  {author} {\bibfnamefont {H.}~\bibnamefont {Niikura}}, \bibinfo {author}
  {\bibfnamefont {H.}~\bibnamefont {P{\'e}pin}}, \bibinfo {author}
  {\bibfnamefont {J.-C.}\ \bibnamefont {Kieffer}}, \bibinfo {author}
  {\bibfnamefont {P.~B.}\ \bibnamefont {Corkum}},\ and\ \bibinfo {author}
  {\bibfnamefont {D.~M.}\ \bibnamefont {Villeneuve}},\ }\bibfield  {title}
  {\bibinfo {title} {Tomographic imaging of molecular orbitals},\ }\href@noop
  {} {\bibfield  {journal} {\bibinfo  {journal} {Nature}\ }\textbf {\bibinfo
  {volume} {432}},\ \bibinfo {pages} {867} (\bibinfo {year}
  {2004})}\BibitemShut {NoStop}%
\bibitem [{\citenamefont {Bian}\ and\ \citenamefont
  {Bandrauk}(2014)}]{probing2014bian}%
  \BibitemOpen
  \bibfield  {author} {\bibinfo {author} {\bibfnamefont {X.-B.}\ \bibnamefont
  {Bian}}\ and\ \bibinfo {author} {\bibfnamefont {A.~D.}\ \bibnamefont
  {Bandrauk}},\ }\bibfield  {title} {\bibinfo {title} {Probing nuclear motion
  by frequency modulation of molecular high-order harmonic generation},\ }\href
  {https://doi.org/10.1103/PhysRevLett.113.193901} {\bibfield  {journal}
  {\bibinfo  {journal} {Phys. Rev. Lett.}\ }\textbf {\bibinfo {volume} {113}},\
  \bibinfo {pages} {193901} (\bibinfo {year} {2014})}\BibitemShut {NoStop}%
\bibitem [{\citenamefont {Dubietis}\ \emph {et~al.}(1992)\citenamefont
  {Dubietis}, \citenamefont {Jonu{\v{s}}auskas},\ and\ \citenamefont
  {Piskarskas}}]{dubietis1992powerful}%
  \BibitemOpen
  \bibfield  {author} {\bibinfo {author} {\bibfnamefont {A.}~\bibnamefont
  {Dubietis}}, \bibinfo {author} {\bibfnamefont {G.}~\bibnamefont
  {Jonu{\v{s}}auskas}},\ and\ \bibinfo {author} {\bibfnamefont
  {A.}~\bibnamefont {Piskarskas}},\ }\bibfield  {title} {\bibinfo {title}
  {Powerful femtosecond pulse generation by chirped and stretched pulse
  parametric amplification in bbo crystal},\ }\href
  {https://doi.org/10.1016/0030-4018(92)90070-8Get} {\bibfield  {journal}
  {\bibinfo  {journal} {Opt. Commun.}\ }\textbf {\bibinfo {volume} {88}},\
  \bibinfo {pages} {437} (\bibinfo {year} {1992})}\BibitemShut {NoStop}%
\bibitem [{\citenamefont {Ishii}\ \emph {et~al.}(2014)\citenamefont {Ishii},
  \citenamefont {Kaneshima}, \citenamefont {Kitano}, \citenamefont {Kanai},
  \citenamefont {Watanabe},\ and\ \citenamefont {Itatani}}]{ishii2014carrier}%
  \BibitemOpen
  \bibfield  {author} {\bibinfo {author} {\bibfnamefont {N.}~\bibnamefont
  {Ishii}}, \bibinfo {author} {\bibfnamefont {K.}~\bibnamefont {Kaneshima}},
  \bibinfo {author} {\bibfnamefont {K.}~\bibnamefont {Kitano}}, \bibinfo
  {author} {\bibfnamefont {T.}~\bibnamefont {Kanai}}, \bibinfo {author}
  {\bibfnamefont {S.}~\bibnamefont {Watanabe}},\ and\ \bibinfo {author}
  {\bibfnamefont {J.}~\bibnamefont {Itatani}},\ }\bibfield  {title} {\bibinfo
  {title} {Carrier-envelope phase-dependent high harmonic generation in the
  water window using few-cycle infrared pulses},\ }\href
  {https://doi.org/10.1038/ncomms4331} {\bibfield  {journal} {\bibinfo
  {journal} {Nat. Commun.}\ }\textbf {\bibinfo {volume} {5}},\ \bibinfo {pages}
  {3331} (\bibinfo {year} {2014})}\BibitemShut {NoStop}%
\bibitem [{\citenamefont {Ndabashimiye}\ \emph {et~al.}(2016)\citenamefont
  {Ndabashimiye}, \citenamefont {Ghimire}, \citenamefont {Wu}, \citenamefont
  {Browne}, \citenamefont {Schafer}, \citenamefont {Gaarde},\ and\
  \citenamefont {Reis}}]{ndabashimiye2016solid}%
  \BibitemOpen
  \bibfield  {author} {\bibinfo {author} {\bibfnamefont {G.}~\bibnamefont
  {Ndabashimiye}}, \bibinfo {author} {\bibfnamefont {S.}~\bibnamefont
  {Ghimire}}, \bibinfo {author} {\bibfnamefont {M.}~\bibnamefont {Wu}},
  \bibinfo {author} {\bibfnamefont {D.~A.}\ \bibnamefont {Browne}}, \bibinfo
  {author} {\bibfnamefont {K.~J.}\ \bibnamefont {Schafer}}, \bibinfo {author}
  {\bibfnamefont {M.~B.}\ \bibnamefont {Gaarde}},\ and\ \bibinfo {author}
  {\bibfnamefont {D.~A.}\ \bibnamefont {Reis}},\ }\bibfield  {title} {\bibinfo
  {title} {Solid-state harmonics beyond the atomic limit},\ }\href
  {https://doi.org/10.1038/nature17660} {\bibfield  {journal} {\bibinfo
  {journal} {Nature}\ }\textbf {\bibinfo {volume} {534}},\ \bibinfo {pages}
  {520} (\bibinfo {year} {2016})}\BibitemShut {NoStop}%
\bibitem [{\citenamefont {You}\ \emph {et~al.}(2017)\citenamefont {You},
  \citenamefont {Wu}, \citenamefont {Yin}, \citenamefont {Chew}, \citenamefont
  {Ren}, \citenamefont {Gholam-Mirzaei}, \citenamefont {Browne}, \citenamefont
  {Chini}, \citenamefont {Chang}, \citenamefont {Schafer}, \citenamefont
  {Gaarde},\ and\ \citenamefont {Ghimire}}]{you2017laser}%
  \BibitemOpen
  \bibfield  {author} {\bibinfo {author} {\bibfnamefont {Y.~S.}\ \bibnamefont
  {You}}, \bibinfo {author} {\bibfnamefont {M.}~\bibnamefont {Wu}}, \bibinfo
  {author} {\bibfnamefont {Y.}~\bibnamefont {Yin}}, \bibinfo {author}
  {\bibfnamefont {A.}~\bibnamefont {Chew}}, \bibinfo {author} {\bibfnamefont
  {X.}~\bibnamefont {Ren}}, \bibinfo {author} {\bibfnamefont {S.}~\bibnamefont
  {Gholam-Mirzaei}}, \bibinfo {author} {\bibfnamefont {D.~A.}\ \bibnamefont
  {Browne}}, \bibinfo {author} {\bibfnamefont {M.}~\bibnamefont {Chini}},
  \bibinfo {author} {\bibfnamefont {Z.}~\bibnamefont {Chang}}, \bibinfo
  {author} {\bibfnamefont {K.~J.}\ \bibnamefont {Schafer}}, \bibinfo {author}
  {\bibfnamefont {M.~B.}\ \bibnamefont {Gaarde}},\ and\ \bibinfo {author}
  {\bibfnamefont {S.}~\bibnamefont {Ghimire}},\ }\bibfield  {title} {\bibinfo
  {title} {Laser waveform control of extreme ultraviolet high harmonics from
  solids},\ }\href {https://doi.org/10.1364/OL.42.001816} {\bibfield  {journal}
  {\bibinfo  {journal} {Opt. Lett.}\ }\textbf {\bibinfo {volume} {42}},\
  \bibinfo {pages} {1816} (\bibinfo {year} {2017})}\BibitemShut {NoStop}%
\bibitem [{\citenamefont {Ghimire}\ \emph {et~al.}(2011)\citenamefont
  {Ghimire}, \citenamefont {DiChiara}, \citenamefont {Sistrunk}, \citenamefont
  {Agostini}, \citenamefont {DiMauro},\ and\ \citenamefont
  {Reis}}]{ghimire2011observation}%
  \BibitemOpen
  \bibfield  {author} {\bibinfo {author} {\bibfnamefont {S.}~\bibnamefont
  {Ghimire}}, \bibinfo {author} {\bibfnamefont {A.~D.}\ \bibnamefont
  {DiChiara}}, \bibinfo {author} {\bibfnamefont {E.}~\bibnamefont {Sistrunk}},
  \bibinfo {author} {\bibfnamefont {P.}~\bibnamefont {Agostini}}, \bibinfo
  {author} {\bibfnamefont {L.~F.}\ \bibnamefont {DiMauro}},\ and\ \bibinfo
  {author} {\bibfnamefont {D.~A.}\ \bibnamefont {Reis}},\ }\bibfield  {title}
  {\bibinfo {title} {Observation of high-order harmonic generation in a bulk
  crystal},\ }\href@noop {} {\bibfield  {journal} {\bibinfo  {journal} {Nature
  Physics}\ }\textbf {\bibinfo {volume} {7}},\ \bibinfo {pages} {138} (\bibinfo
  {year} {2011})}\BibitemShut {NoStop}%
\bibitem [{\citenamefont {Yoshikawa}\ \emph {et~al.}(2017)\citenamefont
  {Yoshikawa}, \citenamefont {Tamaya},\ and\ \citenamefont
  {Tanaka}}]{yoshikawa2017high}%
  \BibitemOpen
  \bibfield  {author} {\bibinfo {author} {\bibfnamefont {N.}~\bibnamefont
  {Yoshikawa}}, \bibinfo {author} {\bibfnamefont {T.}~\bibnamefont {Tamaya}},\
  and\ \bibinfo {author} {\bibfnamefont {K.}~\bibnamefont {Tanaka}},\
  }\bibfield  {title} {\bibinfo {title} {High-harmonic generation in graphene
  enhanced by elliptically polarized light excitation},\ }\href
  {https://www.science.org/doi/10.1126/science.aam8861} {\bibfield  {journal}
  {\bibinfo  {journal} {Science}\ }\textbf {\bibinfo {volume} {356}},\ \bibinfo
  {pages} {736} (\bibinfo {year} {2017})}\BibitemShut {NoStop}%
\bibitem [{\citenamefont {Tancogne-Dejean}\ \emph {et~al.}(2017)\citenamefont
  {Tancogne-Dejean}, \citenamefont {M{\"u}cke}, \citenamefont {K{\"a}rtner},\
  and\ \citenamefont {Rubio}}]{tancogne2017ellipticity}%
  \BibitemOpen
  \bibfield  {author} {\bibinfo {author} {\bibfnamefont {N.}~\bibnamefont
  {Tancogne-Dejean}}, \bibinfo {author} {\bibfnamefont {O.~D.}\ \bibnamefont
  {M{\"u}cke}}, \bibinfo {author} {\bibfnamefont {F.~X.}\ \bibnamefont
  {K{\"a}rtner}},\ and\ \bibinfo {author} {\bibfnamefont {A.}~\bibnamefont
  {Rubio}},\ }\bibfield  {title} {\bibinfo {title} {Ellipticity dependence of
  high-harmonic generation in solids originating from coupled intraband and
  interband dynamics},\ }\href {https://doi.org/10.1038/s41467-017-01768-x}
  {\bibfield  {journal} {\bibinfo  {journal} {Nat. Commun.}\ }\textbf {\bibinfo
  {volume} {8}},\ \bibinfo {pages} {745} (\bibinfo {year} {2017})}\BibitemShut
  {NoStop}%
\bibitem [{\citenamefont {Vampa}\ \emph {et~al.}(2014)\citenamefont {Vampa},
  \citenamefont {McDonald}, \citenamefont {Orlando}, \citenamefont {Klug},
  \citenamefont {Corkum},\ and\ \citenamefont {Brabec}}]{Vampa2014}%
  \BibitemOpen
  \bibfield  {author} {\bibinfo {author} {\bibfnamefont {G.}~\bibnamefont
  {Vampa}}, \bibinfo {author} {\bibfnamefont {C.~R.}\ \bibnamefont {McDonald}},
  \bibinfo {author} {\bibfnamefont {G.}~\bibnamefont {Orlando}}, \bibinfo
  {author} {\bibfnamefont {D.~D.}\ \bibnamefont {Klug}}, \bibinfo {author}
  {\bibfnamefont {P.~B.}\ \bibnamefont {Corkum}},\ and\ \bibinfo {author}
  {\bibfnamefont {T.}~\bibnamefont {Brabec}},\ }\bibfield  {title} {\bibinfo
  {title} {Theoretical analysis of high-harmonic generation in solids},\ }\href
  {https://doi.org/10.1103/PhysRevLett.113.073901} {\bibfield  {journal}
  {\bibinfo  {journal} {Phys. Rev. Lett.}\ }\textbf {\bibinfo {volume} {113}},\
  \bibinfo {pages} {073901} (\bibinfo {year} {2014})}\BibitemShut {NoStop}%
\bibitem [{\citenamefont {Vampa}\ \emph {et~al.}(2015)\citenamefont {Vampa},
  \citenamefont {McDonald}, \citenamefont {Orlando}, \citenamefont {Corkum},\
  and\ \citenamefont {Brabec}}]{vampa2015semiclassical}%
  \BibitemOpen
  \bibfield  {author} {\bibinfo {author} {\bibfnamefont {G.}~\bibnamefont
  {Vampa}}, \bibinfo {author} {\bibfnamefont {C.~R.}\ \bibnamefont {McDonald}},
  \bibinfo {author} {\bibfnamefont {G.}~\bibnamefont {Orlando}}, \bibinfo
  {author} {\bibfnamefont {P.~B.}\ \bibnamefont {Corkum}},\ and\ \bibinfo
  {author} {\bibfnamefont {T.}~\bibnamefont {Brabec}},\ }\bibfield  {title}
  {\bibinfo {title} {Semiclassical analysis of high harmonic generation in bulk
  crystals},\ }\href {https://doi.org/10.1103/PhysRevB.91.064302} {\bibfield
  {journal} {\bibinfo  {journal} {Phys. Rev. B}\ }\textbf {\bibinfo {volume}
  {91}},\ \bibinfo {pages} {064302} (\bibinfo {year} {2015})}\BibitemShut
  {NoStop}%
\bibitem [{\citenamefont {Han}\ \emph {et~al.}(2016)\citenamefont {Han},
  \citenamefont {Kim}, \citenamefont {Kim}, \citenamefont {Kim}, \citenamefont
  {Kim}, \citenamefont {Park},\ and\ \citenamefont {Kim}}]{han2016high}%
  \BibitemOpen
  \bibfield  {author} {\bibinfo {author} {\bibfnamefont {S.}~\bibnamefont
  {Han}}, \bibinfo {author} {\bibfnamefont {H.}~\bibnamefont {Kim}}, \bibinfo
  {author} {\bibfnamefont {Y.~W.}\ \bibnamefont {Kim}}, \bibinfo {author}
  {\bibfnamefont {Y.-J.}\ \bibnamefont {Kim}}, \bibinfo {author} {\bibfnamefont
  {S.}~\bibnamefont {Kim}}, \bibinfo {author} {\bibfnamefont {I.-Y.}\
  \bibnamefont {Park}},\ and\ \bibinfo {author} {\bibfnamefont {S.-W.}\
  \bibnamefont {Kim}},\ }\bibfield  {title} {\bibinfo {title} {High-harmonic
  generation by field enhanced femtosecond pulses in metal-sapphire
  nanostructure},\ }\href@noop {} {\bibfield  {journal} {\bibinfo  {journal}
  {Nature communications}\ }\textbf {\bibinfo {volume} {7}},\ \bibinfo {pages}
  {13105} (\bibinfo {year} {2016})}\BibitemShut {NoStop}%
\bibitem [{\citenamefont {Vampa}\ \emph {et~al.}(2017)\citenamefont {Vampa},
  \citenamefont {Ghamsari}, \citenamefont {Siadat~Mousavi}, \citenamefont
  {Hammond}, \citenamefont {Olivieri}, \citenamefont {Lisicka-Skrek},
  \citenamefont {Naumov}, \citenamefont {Villeneuve}, \citenamefont {Staudte},
  \citenamefont {Berini} \emph {et~al.}}]{vampa2017plasmon}%
  \BibitemOpen
  \bibfield  {author} {\bibinfo {author} {\bibfnamefont {G.}~\bibnamefont
  {Vampa}}, \bibinfo {author} {\bibfnamefont {B.}~\bibnamefont {Ghamsari}},
  \bibinfo {author} {\bibfnamefont {S.}~\bibnamefont {Siadat~Mousavi}},
  \bibinfo {author} {\bibfnamefont {T.}~\bibnamefont {Hammond}}, \bibinfo
  {author} {\bibfnamefont {A.}~\bibnamefont {Olivieri}}, \bibinfo {author}
  {\bibfnamefont {E.}~\bibnamefont {Lisicka-Skrek}}, \bibinfo {author}
  {\bibfnamefont {A.~Y.}\ \bibnamefont {Naumov}}, \bibinfo {author}
  {\bibfnamefont {D.}~\bibnamefont {Villeneuve}}, \bibinfo {author}
  {\bibfnamefont {A.}~\bibnamefont {Staudte}}, \bibinfo {author} {\bibfnamefont
  {P.}~\bibnamefont {Berini}}, \emph {et~al.},\ }\bibfield  {title} {\bibinfo
  {title} {Plasmon-enhanced high-harmonic generation from silicon},\
  }\href@noop {} {\bibfield  {journal} {\bibinfo  {journal} {Nature Physics}\
  }\textbf {\bibinfo {volume} {13}},\ \bibinfo {pages} {659} (\bibinfo {year}
  {2017})}\BibitemShut {NoStop}%
\bibitem [{\citenamefont {Liu}\ \emph {et~al.}(2018)\citenamefont {Liu},
  \citenamefont {Guo}, \citenamefont {Vampa}, \citenamefont {Zhang},
  \citenamefont {Sarmiento}, \citenamefont {Xiao}, \citenamefont {Bucksbaum},
  \citenamefont {Vu{\v{c}}kovi{\'c}}, \citenamefont {Fan},\ and\ \citenamefont
  {Reis}}]{liu2018enhanced}%
  \BibitemOpen
  \bibfield  {author} {\bibinfo {author} {\bibfnamefont {H.}~\bibnamefont
  {Liu}}, \bibinfo {author} {\bibfnamefont {C.}~\bibnamefont {Guo}}, \bibinfo
  {author} {\bibfnamefont {G.}~\bibnamefont {Vampa}}, \bibinfo {author}
  {\bibfnamefont {J.~L.}\ \bibnamefont {Zhang}}, \bibinfo {author}
  {\bibfnamefont {T.}~\bibnamefont {Sarmiento}}, \bibinfo {author}
  {\bibfnamefont {M.}~\bibnamefont {Xiao}}, \bibinfo {author} {\bibfnamefont
  {P.~H.}\ \bibnamefont {Bucksbaum}}, \bibinfo {author} {\bibfnamefont
  {J.}~\bibnamefont {Vu{\v{c}}kovi{\'c}}}, \bibinfo {author} {\bibfnamefont
  {S.}~\bibnamefont {Fan}},\ and\ \bibinfo {author} {\bibfnamefont {D.~A.}\
  \bibnamefont {Reis}},\ }\bibfield  {title} {\bibinfo {title} {Enhanced
  high-harmonic generation from an all-dielectric metasurface},\ }\href@noop {}
  {\bibfield  {journal} {\bibinfo  {journal} {Nature Physics}\ }\textbf
  {\bibinfo {volume} {14}},\ \bibinfo {pages} {1006} (\bibinfo {year}
  {2018})}\BibitemShut {NoStop}%
\bibitem [{\citenamefont {Sivis}\ \emph {et~al.}(2017)\citenamefont {Sivis},
  \citenamefont {Taucer}, \citenamefont {Vampa}, \citenamefont {Johnston},
  \citenamefont {Staudte}, \citenamefont {Naumov}, \citenamefont {Villeneuve},
  \citenamefont {Ropers},\ and\ \citenamefont {Corkum}}]{sivis2017tailored}%
  \BibitemOpen
  \bibfield  {author} {\bibinfo {author} {\bibfnamefont {M.}~\bibnamefont
  {Sivis}}, \bibinfo {author} {\bibfnamefont {M.}~\bibnamefont {Taucer}},
  \bibinfo {author} {\bibfnamefont {G.}~\bibnamefont {Vampa}}, \bibinfo
  {author} {\bibfnamefont {K.}~\bibnamefont {Johnston}}, \bibinfo {author}
  {\bibfnamefont {A.}~\bibnamefont {Staudte}}, \bibinfo {author} {\bibfnamefont
  {A.~Y.}\ \bibnamefont {Naumov}}, \bibinfo {author} {\bibfnamefont
  {D.}~\bibnamefont {Villeneuve}}, \bibinfo {author} {\bibfnamefont
  {C.}~\bibnamefont {Ropers}},\ and\ \bibinfo {author} {\bibfnamefont
  {P.}~\bibnamefont {Corkum}},\ }\bibfield  {title} {\bibinfo {title} {Tailored
  semiconductors for high-harmonic optoelectronics},\ }\href@noop {} {\bibfield
   {journal} {\bibinfo  {journal} {Science}\ }\textbf {\bibinfo {volume}
  {357}},\ \bibinfo {pages} {303} (\bibinfo {year} {2017})}\BibitemShut
  {NoStop}%
\bibitem [{\citenamefont {Lou}\ \emph {et~al.}(2020)\citenamefont {Lou},
  \citenamefont {Zheng}, \citenamefont {Liu}, \citenamefont {Zhang},
  \citenamefont {Ge}, \citenamefont {Li}, \citenamefont {Wang}, \citenamefont
  {Zeng}, \citenamefont {Li},\ and\ \citenamefont {Xu}}]{lou2020ellipticity}%
  \BibitemOpen
  \bibfield  {author} {\bibinfo {author} {\bibfnamefont {Z.}~\bibnamefont
  {Lou}}, \bibinfo {author} {\bibfnamefont {Y.}~\bibnamefont {Zheng}}, \bibinfo
  {author} {\bibfnamefont {C.}~\bibnamefont {Liu}}, \bibinfo {author}
  {\bibfnamefont {L.}~\bibnamefont {Zhang}}, \bibinfo {author} {\bibfnamefont
  {X.}~\bibnamefont {Ge}}, \bibinfo {author} {\bibfnamefont {Y.}~\bibnamefont
  {Li}}, \bibinfo {author} {\bibfnamefont {J.}~\bibnamefont {Wang}}, \bibinfo
  {author} {\bibfnamefont {Z.}~\bibnamefont {Zeng}}, \bibinfo {author}
  {\bibfnamefont {R.}~\bibnamefont {Li}},\ and\ \bibinfo {author}
  {\bibfnamefont {Z.}~\bibnamefont {Xu}},\ }\bibfield  {title} {\bibinfo
  {title} {Ellipticity dependence of nonperturbative harmonic generation in
  few-layer mos2},\ }\href@noop {} {\bibfield  {journal} {\bibinfo  {journal}
  {Optics Communications}\ }\textbf {\bibinfo {volume} {469}},\ \bibinfo
  {pages} {125769} (\bibinfo {year} {2020})}\BibitemShut {NoStop}%
\bibitem [{\citenamefont {Liu}\ \emph {et~al.}(2017)\citenamefont {Liu},
  \citenamefont {Li}, \citenamefont {You}, \citenamefont {Ghimire},
  \citenamefont {Heinz},\ and\ \citenamefont {Reis}}]{liu2017high}%
  \BibitemOpen
  \bibfield  {author} {\bibinfo {author} {\bibfnamefont {H.}~\bibnamefont
  {Liu}}, \bibinfo {author} {\bibfnamefont {Y.}~\bibnamefont {Li}}, \bibinfo
  {author} {\bibfnamefont {Y.~S.}\ \bibnamefont {You}}, \bibinfo {author}
  {\bibfnamefont {S.}~\bibnamefont {Ghimire}}, \bibinfo {author} {\bibfnamefont
  {T.~F.}\ \bibnamefont {Heinz}},\ and\ \bibinfo {author} {\bibfnamefont
  {D.~A.}\ \bibnamefont {Reis}},\ }\bibfield  {title} {\bibinfo {title}
  {High-harmonic generation from an atomically thin semiconductor},\
  }\href@noop {} {\bibfield  {journal} {\bibinfo  {journal} {Nature Physics}\
  }\textbf {\bibinfo {volume} {13}},\ \bibinfo {pages} {262} (\bibinfo {year}
  {2017})}\BibitemShut {NoStop}%
\bibitem [{\citenamefont {McDonald}\ \emph {et~al.}(2017)\citenamefont
  {McDonald}, \citenamefont {Amin}, \citenamefont {Aalmalki},\ and\
  \citenamefont {Brabec}}]{mcdonald2017enhancing}%
  \BibitemOpen
  \bibfield  {author} {\bibinfo {author} {\bibfnamefont {C.}~\bibnamefont
  {McDonald}}, \bibinfo {author} {\bibfnamefont {K.}~\bibnamefont {Amin}},
  \bibinfo {author} {\bibfnamefont {S.}~\bibnamefont {Aalmalki}},\ and\
  \bibinfo {author} {\bibfnamefont {T.}~\bibnamefont {Brabec}},\ }\bibfield
  {title} {\bibinfo {title} {Enhancing high harmonic output in solids through
  quantum confinement},\ }\href@noop {} {\bibfield  {journal} {\bibinfo
  {journal} {Physical Review Letters}\ }\textbf {\bibinfo {volume} {119}},\
  \bibinfo {pages} {183902} (\bibinfo {year} {2017})}\BibitemShut {NoStop}%
\bibitem [{\citenamefont {Le~Breton}\ \emph {et~al.}(2018)\citenamefont
  {Le~Breton}, \citenamefont {Rubio},\ and\ \citenamefont
  {Tancogne-Dejean}}]{le2018high}%
  \BibitemOpen
  \bibfield  {author} {\bibinfo {author} {\bibfnamefont {G.}~\bibnamefont
  {Le~Breton}}, \bibinfo {author} {\bibfnamefont {A.}~\bibnamefont {Rubio}},\
  and\ \bibinfo {author} {\bibfnamefont {N.}~\bibnamefont {Tancogne-Dejean}},\
  }\bibfield  {title} {\bibinfo {title} {High-harmonic generation from
  few-layer hexagonal boron nitride: Evolution from monolayer to bulk
  response},\ }\href@noop {} {\bibfield  {journal} {\bibinfo  {journal}
  {Physical Review B}\ }\textbf {\bibinfo {volume} {98}},\ \bibinfo {pages}
  {165308} (\bibinfo {year} {2018})}\BibitemShut {NoStop}%
\bibitem [{\citenamefont {Alonso~Calafell}\ \emph {et~al.}(2021)\citenamefont
  {Alonso~Calafell}, \citenamefont {Rozema}, \citenamefont {Alcaraz~Iranzo},
  \citenamefont {Trenti}, \citenamefont {Jenke}, \citenamefont {Cox},
  \citenamefont {Kumar}, \citenamefont {Bieliaiev}, \citenamefont {Nanot},
  \citenamefont {Peng} \emph {et~al.}}]{alonso2021giant}%
  \BibitemOpen
  \bibfield  {author} {\bibinfo {author} {\bibfnamefont {I.}~\bibnamefont
  {Alonso~Calafell}}, \bibinfo {author} {\bibfnamefont {L.~A.}\ \bibnamefont
  {Rozema}}, \bibinfo {author} {\bibfnamefont {D.}~\bibnamefont
  {Alcaraz~Iranzo}}, \bibinfo {author} {\bibfnamefont {A.}~\bibnamefont
  {Trenti}}, \bibinfo {author} {\bibfnamefont {P.~K.}\ \bibnamefont {Jenke}},
  \bibinfo {author} {\bibfnamefont {J.~D.}\ \bibnamefont {Cox}}, \bibinfo
  {author} {\bibfnamefont {A.}~\bibnamefont {Kumar}}, \bibinfo {author}
  {\bibfnamefont {H.}~\bibnamefont {Bieliaiev}}, \bibinfo {author}
  {\bibfnamefont {S.}~\bibnamefont {Nanot}}, \bibinfo {author} {\bibfnamefont
  {C.}~\bibnamefont {Peng}}, \emph {et~al.},\ }\bibfield  {title} {\bibinfo
  {title} {Giant enhancement of third-harmonic generation in graphene--metal
  heterostructures},\ }\href@noop {} {\bibfield  {journal} {\bibinfo  {journal}
  {Nature Nanotechnology}\ }\textbf {\bibinfo {volume} {16}},\ \bibinfo {pages}
  {318} (\bibinfo {year} {2021})}\BibitemShut {NoStop}%
\bibitem [{\citenamefont {Qin}\ and\ \citenamefont
  {Chen}(2018)}]{qin2018strain}%
  \BibitemOpen
  \bibfield  {author} {\bibinfo {author} {\bibfnamefont {R.}~\bibnamefont
  {Qin}}\ and\ \bibinfo {author} {\bibfnamefont {Z.-Y.}\ \bibnamefont {Chen}},\
  }\bibfield  {title} {\bibinfo {title} {Strain-controlled high harmonic
  generation with dirac fermions in silicene},\ }\href@noop {} {\bibfield
  {journal} {\bibinfo  {journal} {Nanoscale}\ }\textbf {\bibinfo {volume}
  {10}},\ \bibinfo {pages} {22593} (\bibinfo {year} {2018})}\BibitemShut
  {NoStop}%
\bibitem [{\citenamefont {Shao}\ \emph {et~al.}(2019)\citenamefont {Shao},
  \citenamefont {Xu}, \citenamefont {Huang},\ and\ \citenamefont
  {Bian}}]{strain2019shao}%
  \BibitemOpen
  \bibfield  {author} {\bibinfo {author} {\bibfnamefont {T.-J.}\ \bibnamefont
  {Shao}}, \bibinfo {author} {\bibfnamefont {Y.}~\bibnamefont {Xu}}, \bibinfo
  {author} {\bibfnamefont {X.-H.}\ \bibnamefont {Huang}},\ and\ \bibinfo
  {author} {\bibfnamefont {X.-B.}\ \bibnamefont {Bian}},\ }\bibfield  {title}
  {\bibinfo {title} {Strain effects on high-order harmonic generation in
  solids},\ }\href {https://doi.org/10.1103/PhysRevA.99.013432} {\bibfield
  {journal} {\bibinfo  {journal} {Phys. Rev. A}\ }\textbf {\bibinfo {volume}
  {99}},\ \bibinfo {pages} {013432} (\bibinfo {year} {2019})}\BibitemShut
  {NoStop}%
\bibitem [{\citenamefont {Tamaya}\ \emph {et~al.}(2022)\citenamefont {Tamaya},
  \citenamefont {Akiyama},\ and\ \citenamefont {Kato}}]{tamaya2022shear}%
  \BibitemOpen
  \bibfield  {author} {\bibinfo {author} {\bibfnamefont {T.}~\bibnamefont
  {Tamaya}}, \bibinfo {author} {\bibfnamefont {H.}~\bibnamefont {Akiyama}},\
  and\ \bibinfo {author} {\bibfnamefont {T.}~\bibnamefont {Kato}},\ }\bibfield
  {title} {\bibinfo {title} {Shear-strain controlled high-harmonic generation
  in graphene},\ }\href@noop {} {\bibfield  {journal} {\bibinfo  {journal}
  {arXiv preprint arXiv:2212.10111}\ } (\bibinfo {year} {2022})}\BibitemShut
  {NoStop}%
\bibitem [{\citenamefont {Nefedova}\ \emph {et~al.}(2021)\citenamefont
  {Nefedova}, \citenamefont {Fr{\"o}hlich}, \citenamefont {Navarrete},
  \citenamefont {Tancogne-Dejean}, \citenamefont {Franz}, \citenamefont
  {Hamdou}, \citenamefont {Kaassamani}, \citenamefont {Gauthier}, \citenamefont
  {Nicolas}, \citenamefont {Jargot} \emph {et~al.}}]{nefedova2021enhanced}%
  \BibitemOpen
  \bibfield  {author} {\bibinfo {author} {\bibfnamefont {V.}~\bibnamefont
  {Nefedova}}, \bibinfo {author} {\bibfnamefont {S.}~\bibnamefont
  {Fr{\"o}hlich}}, \bibinfo {author} {\bibfnamefont {F.}~\bibnamefont
  {Navarrete}}, \bibinfo {author} {\bibfnamefont {N.}~\bibnamefont
  {Tancogne-Dejean}}, \bibinfo {author} {\bibfnamefont {D.}~\bibnamefont
  {Franz}}, \bibinfo {author} {\bibfnamefont {A.}~\bibnamefont {Hamdou}},
  \bibinfo {author} {\bibfnamefont {S.}~\bibnamefont {Kaassamani}}, \bibinfo
  {author} {\bibfnamefont {D.}~\bibnamefont {Gauthier}}, \bibinfo {author}
  {\bibfnamefont {R.}~\bibnamefont {Nicolas}}, \bibinfo {author} {\bibfnamefont
  {G.}~\bibnamefont {Jargot}}, \emph {et~al.},\ }\bibfield  {title} {\bibinfo
  {title} {Enhanced extreme ultraviolet high-harmonic generation from
  chromium-doped magnesium oxide},\ }\href@noop {} {\bibfield  {journal}
  {\bibinfo  {journal} {Applied Physics Letters}\ }\textbf {\bibinfo {volume}
  {118}},\ \bibinfo {pages} {201103} (\bibinfo {year} {2021})}\BibitemShut
  {NoStop}%
\bibitem [{\citenamefont {Wang}\ \emph {et~al.}(2017)\citenamefont {Wang},
  \citenamefont {Park}, \citenamefont {Lai}, \citenamefont {Xu}, \citenamefont
  {Blaga}, \citenamefont {Yang}, \citenamefont {Agostini},\ and\ \citenamefont
  {DiMauro}}]{wang2017roles}%
  \BibitemOpen
  \bibfield  {author} {\bibinfo {author} {\bibfnamefont {Z.}~\bibnamefont
  {Wang}}, \bibinfo {author} {\bibfnamefont {H.}~\bibnamefont {Park}}, \bibinfo
  {author} {\bibfnamefont {Y.~H.}\ \bibnamefont {Lai}}, \bibinfo {author}
  {\bibfnamefont {J.}~\bibnamefont {Xu}}, \bibinfo {author} {\bibfnamefont
  {C.~I.}\ \bibnamefont {Blaga}}, \bibinfo {author} {\bibfnamefont
  {F.}~\bibnamefont {Yang}}, \bibinfo {author} {\bibfnamefont {P.}~\bibnamefont
  {Agostini}},\ and\ \bibinfo {author} {\bibfnamefont {L.~F.}\ \bibnamefont
  {DiMauro}},\ }\bibfield  {title} {\bibinfo {title} {The roles of
  photo-carrier doping and driving wavelength in high harmonic generation from
  a semiconductor},\ }\href@noop {} {\bibfield  {journal} {\bibinfo  {journal}
  {Nature communications}\ }\textbf {\bibinfo {volume} {8}},\ \bibinfo {pages}
  {1686} (\bibinfo {year} {2017})}\BibitemShut {NoStop}%
\bibitem [{\citenamefont {Huang}\ \emph {et~al.}(2017)\citenamefont {Huang},
  \citenamefont {Zhu}, \citenamefont {Li}, \citenamefont {Liu}, \citenamefont
  {Lan},\ and\ \citenamefont {Lu}}]{huang2017high}%
  \BibitemOpen
  \bibfield  {author} {\bibinfo {author} {\bibfnamefont {T.}~\bibnamefont
  {Huang}}, \bibinfo {author} {\bibfnamefont {X.}~\bibnamefont {Zhu}}, \bibinfo
  {author} {\bibfnamefont {L.}~\bibnamefont {Li}}, \bibinfo {author}
  {\bibfnamefont {X.}~\bibnamefont {Liu}}, \bibinfo {author} {\bibfnamefont
  {P.}~\bibnamefont {Lan}},\ and\ \bibinfo {author} {\bibfnamefont
  {P.}~\bibnamefont {Lu}},\ }\bibfield  {title} {\bibinfo {title}
  {High-order-harmonic generation of a doped semiconductor},\ }\href@noop {}
  {\bibfield  {journal} {\bibinfo  {journal} {Physical Review A}\ }\textbf
  {\bibinfo {volume} {96}},\ \bibinfo {pages} {043425} (\bibinfo {year}
  {2017})}\BibitemShut {NoStop}%
\bibitem [{\citenamefont {Yu}\ \emph {et~al.}(2019)\citenamefont {Yu},
  \citenamefont {Hansen},\ and\ \citenamefont {Madsen}}]{yu2019enhanced}%
  \BibitemOpen
  \bibfield  {author} {\bibinfo {author} {\bibfnamefont {C.}~\bibnamefont
  {Yu}}, \bibinfo {author} {\bibfnamefont {K.~K.}\ \bibnamefont {Hansen}},\
  and\ \bibinfo {author} {\bibfnamefont {L.~B.}\ \bibnamefont {Madsen}},\
  }\bibfield  {title} {\bibinfo {title} {Enhanced high-order harmonic
  generation in donor-doped band-gap materials},\ }\href@noop {} {\bibfield
  {journal} {\bibinfo  {journal} {Physical Review A}\ }\textbf {\bibinfo
  {volume} {99}},\ \bibinfo {pages} {013435} (\bibinfo {year}
  {2019})}\BibitemShut {NoStop}%
\bibitem [{\citenamefont {Jia}\ \emph {et~al.}(2019)\citenamefont {Jia},
  \citenamefont {Zhang}, \citenamefont {Yang}, \citenamefont {Si},
  \citenamefont {Zhang},\ and\ \citenamefont {Liu}}]{high2019jia}%
  \BibitemOpen
  \bibfield  {author} {\bibinfo {author} {\bibfnamefont {L.}~\bibnamefont
  {Jia}}, \bibinfo {author} {\bibfnamefont {Z.}~\bibnamefont {Zhang}}, \bibinfo
  {author} {\bibfnamefont {D.~Z.}\ \bibnamefont {Yang}}, \bibinfo {author}
  {\bibfnamefont {M.~S.}\ \bibnamefont {Si}}, \bibinfo {author} {\bibfnamefont
  {G.~P.}\ \bibnamefont {Zhang}},\ and\ \bibinfo {author} {\bibfnamefont
  {Y.~S.}\ \bibnamefont {Liu}},\ }\bibfield  {title} {\bibinfo {title} {High
  harmonic generation in magnetically-doped topological insulators},\ }\href
  {https://doi.org/10.1103/PhysRevB.100.125144} {\bibfield  {journal} {\bibinfo
   {journal} {Phys. Rev. B}\ }\textbf {\bibinfo {volume} {100}},\ \bibinfo
  {pages} {125144} (\bibinfo {year} {2019})}\BibitemShut {NoStop}%
\bibitem [{\citenamefont {Mrudul}\ \emph {et~al.}(2020)\citenamefont {Mrudul},
  \citenamefont {Tancogne-Dejean}, \citenamefont {Rubio},\ and\ \citenamefont
  {Dixit}}]{mrudul2020high}%
  \BibitemOpen
  \bibfield  {author} {\bibinfo {author} {\bibfnamefont {M.}~\bibnamefont
  {Mrudul}}, \bibinfo {author} {\bibfnamefont {N.}~\bibnamefont
  {Tancogne-Dejean}}, \bibinfo {author} {\bibfnamefont {A.}~\bibnamefont
  {Rubio}},\ and\ \bibinfo {author} {\bibfnamefont {G.}~\bibnamefont {Dixit}},\
  }\bibfield  {title} {\bibinfo {title} {High-harmonic generation from
  spin-polarised defects in solids},\ }\href@noop {} {\bibfield  {journal}
  {\bibinfo  {journal} {npj Computational Materials}\ }\textbf {\bibinfo
  {volume} {6}},\ \bibinfo {pages} {10} (\bibinfo {year} {2020})}\BibitemShut
  {NoStop}%
\bibitem [{\citenamefont {Pattanayak}\ \emph {et~al.}(2020)\citenamefont
  {Pattanayak}, \citenamefont {S.},\ and\ \citenamefont
  {Dixit}}]{pattanayak2020influence}%
  \BibitemOpen
  \bibfield  {author} {\bibinfo {author} {\bibfnamefont {A.}~\bibnamefont
  {Pattanayak}}, \bibinfo {author} {\bibfnamefont {M.~M.}\ \bibnamefont {S.}},\
  and\ \bibinfo {author} {\bibfnamefont {G.}~\bibnamefont {Dixit}},\ }\bibfield
   {title} {\bibinfo {title} {Influence of vacancy defects in solid high-order
  harmonic generation},\ }\href {https://doi.org/10.1103/PhysRevA.101.013404}
  {\bibfield  {journal} {\bibinfo  {journal} {Phys. Rev. A}\ }\textbf {\bibinfo
  {volume} {101}},\ \bibinfo {pages} {013404} (\bibinfo {year}
  {2020})}\BibitemShut {NoStop}%
\bibitem [{\citenamefont {Zhao}\ \emph {et~al.}(2021)\citenamefont {Zhao},
  \citenamefont {Wang}, \citenamefont {Ding},\ and\ \citenamefont
  {Du}}]{zhao2021impact}%
  \BibitemOpen
  \bibfield  {author} {\bibinfo {author} {\bibfnamefont {Y.-P.}\ \bibnamefont
  {Zhao}}, \bibinfo {author} {\bibfnamefont {G.}~\bibnamefont {Wang}}, \bibinfo
  {author} {\bibfnamefont {S.-J.}\ \bibnamefont {Ding}},\ and\ \bibinfo
  {author} {\bibfnamefont {T.-Y.}\ \bibnamefont {Du}},\ }\bibfield  {title}
  {\bibinfo {title} {Impact of donor and acceptor dopants in high-harmonic
  generation spectra of solids},\ }\href@noop {} {\bibfield  {journal}
  {\bibinfo  {journal} {JOSA B}\ }\textbf {\bibinfo {volume} {38}},\ \bibinfo
  {pages} {2223} (\bibinfo {year} {2021})}\BibitemShut {NoStop}%
\bibitem [{\citenamefont {Wu}\ \emph {et~al.}(2015)\citenamefont {Wu},
  \citenamefont {Ghimire}, \citenamefont {Reis}, \citenamefont {Schafer},\ and\
  \citenamefont {Gaarde}}]{wu2015high}%
  \BibitemOpen
  \bibfield  {author} {\bibinfo {author} {\bibfnamefont {M.}~\bibnamefont
  {Wu}}, \bibinfo {author} {\bibfnamefont {S.}~\bibnamefont {Ghimire}},
  \bibinfo {author} {\bibfnamefont {D.~A.}\ \bibnamefont {Reis}}, \bibinfo
  {author} {\bibfnamefont {K.~J.}\ \bibnamefont {Schafer}},\ and\ \bibinfo
  {author} {\bibfnamefont {M.~B.}\ \bibnamefont {Gaarde}},\ }\bibfield  {title}
  {\bibinfo {title} {High-harmonic generation from bloch electrons in solids},\
  }\href {https://doi.org/10.1103/PhysRevA.91.043839} {\bibfield  {journal}
  {\bibinfo  {journal} {Phys. Rev. A}\ }\textbf {\bibinfo {volume} {91}},\
  \bibinfo {pages} {043839} (\bibinfo {year} {2015})}\BibitemShut {NoStop}%
\bibitem [{\citenamefont {Guan}\ \emph {et~al.}(2016)\citenamefont {Guan},
  \citenamefont {Zhou},\ and\ \citenamefont {Bian}}]{guan2016high}%
  \BibitemOpen
  \bibfield  {author} {\bibinfo {author} {\bibfnamefont {Z.}~\bibnamefont
  {Guan}}, \bibinfo {author} {\bibfnamefont {X.-X.}\ \bibnamefont {Zhou}},\
  and\ \bibinfo {author} {\bibfnamefont {X.-B.}\ \bibnamefont {Bian}},\
  }\bibfield  {title} {\bibinfo {title} {High-order-harmonic generation from
  periodic potentials driven by few-cycle laser pulses},\ }\href
  {https://doi.org/10.1103/PhysRevA.93.033852} {\bibfield  {journal} {\bibinfo
  {journal} {Phys. Rev. A}\ }\textbf {\bibinfo {volume} {93}},\ \bibinfo
  {pages} {033852} (\bibinfo {year} {2016})}\BibitemShut {NoStop}%
\bibitem [{\citenamefont {Wang}\ and\ \citenamefont
  {Bian}(2021)}]{Wang2021Model}%
  \BibitemOpen
  \bibfield  {author} {\bibinfo {author} {\bibfnamefont {X.-Q.}\ \bibnamefont
  {Wang}}\ and\ \bibinfo {author} {\bibfnamefont {X.-B.}\ \bibnamefont
  {Bian}},\ }\bibfield  {title} {\bibinfo {title} {Model-potential method for
  high-order harmonic generation in monolayer graphene},\ }\href
  {https://doi.org/10.1103/PhysRevA.103.053106} {\bibfield  {journal} {\bibinfo
   {journal} {Phys. Rev. A}\ }\textbf {\bibinfo {volume} {103}},\ \bibinfo
  {pages} {053106} (\bibinfo {year} {2021})}\BibitemShut {NoStop}%
\bibitem [{\citenamefont {Hansen}\ \emph {et~al.}(2018)\citenamefont {Hansen},
  \citenamefont {Bauer},\ and\ \citenamefont
  {Madsen}}]{hansen2017finite-system}%
  \BibitemOpen
  \bibfield  {author} {\bibinfo {author} {\bibfnamefont {K.~K.}\ \bibnamefont
  {Hansen}}, \bibinfo {author} {\bibfnamefont {D.}~\bibnamefont {Bauer}},\ and\
  \bibinfo {author} {\bibfnamefont {L.~B.}\ \bibnamefont {Madsen}},\ }\bibfield
   {title} {\bibinfo {title} {Finite-system effects on high-order harmonic
  generation: From atoms to solids},\ }\href
  {https://doi.org/10.1103/PhysRevA.97.043424} {\bibfield  {journal} {\bibinfo
  {journal} {Phys. Rev. A}\ }\textbf {\bibinfo {volume} {97}},\ \bibinfo
  {pages} {043424} (\bibinfo {year} {2018})}\BibitemShut {NoStop}%
\bibitem [{\citenamefont {Hansen}\ \emph {et~al.}(2017)\citenamefont {Hansen},
  \citenamefont {Deffge},\ and\ \citenamefont {Bauer}}]{hansen2017high-order}%
  \BibitemOpen
  \bibfield  {author} {\bibinfo {author} {\bibfnamefont {K.~K.}\ \bibnamefont
  {Hansen}}, \bibinfo {author} {\bibfnamefont {T.}~\bibnamefont {Deffge}},\
  and\ \bibinfo {author} {\bibfnamefont {D.}~\bibnamefont {Bauer}},\ }\bibfield
   {title} {\bibinfo {title} {High-order harmonic generation in solid slabs
  beyond the single-active-electron approximation},\ }\href
  {https://doi.org/10.1103/PhysRevA.96.053418} {\bibfield  {journal} {\bibinfo
  {journal} {Phys. Rev. A}\ }\textbf {\bibinfo {volume} {96}},\ \bibinfo
  {pages} {053418} (\bibinfo {year} {2017})}\BibitemShut {NoStop}%
\bibitem [{\citenamefont {Feit}\ \emph {et~al.}(1982)\citenamefont {Feit},
  \citenamefont {Fleck},\ and\ \citenamefont {Steiger}}]{feit1982solution}%
  \BibitemOpen
  \bibfield  {author} {\bibinfo {author} {\bibfnamefont {M.~D.}\ \bibnamefont
  {Feit}}, \bibinfo {author} {\bibfnamefont {J.~A.}\ \bibnamefont {Fleck}},\
  and\ \bibinfo {author} {\bibfnamefont {A.}~\bibnamefont {Steiger}},\
  }\bibfield  {title} {\bibinfo {title} {Solution of the schrödinger equation
  by a spectral method},\ }\href {https://doi.org/10.1016/0021-9991(82)90091-2}
  {\bibfield  {journal} {\bibinfo  {journal} {J. Comput. Phys.}\ }\textbf
  {\bibinfo {volume} {47}},\ \bibinfo {pages} {412} (\bibinfo {year}
  {1982})}\BibitemShut {NoStop}%
\bibitem [{\citenamefont {Du}\ and\ \citenamefont
  {Bian}(2017)}]{du2017quasi-classical}%
  \BibitemOpen
  \bibfield  {author} {\bibinfo {author} {\bibfnamefont {T.-Y.}\ \bibnamefont
  {Du}}\ and\ \bibinfo {author} {\bibfnamefont {X.-B.}\ \bibnamefont {Bian}},\
  }\bibfield  {title} {\bibinfo {title} {Quasi-classical analysis of the
  dynamics of the high-order harmonic generation from solids},\ }\href
  {https://doi.org/10.1364/OE.25.000151} {\bibfield  {journal} {\bibinfo
  {journal} {Opt. Express}\ }\textbf {\bibinfo {volume} {25}},\ \bibinfo
  {pages} {151} (\bibinfo {year} {2017})}\BibitemShut {NoStop}%
\bibitem [{\citenamefont {Jia}\ \emph {et~al.}(2017)\citenamefont {Jia},
  \citenamefont {Huang},\ and\ \citenamefont {Bian}}]{jia2017nonadiabatic}%
  \BibitemOpen
  \bibfield  {author} {\bibinfo {author} {\bibfnamefont {G.-R.}\ \bibnamefont
  {Jia}}, \bibinfo {author} {\bibfnamefont {X.-H.}\ \bibnamefont {Huang}},\
  and\ \bibinfo {author} {\bibfnamefont {X.-B.}\ \bibnamefont {Bian}},\
  }\bibfield  {title} {\bibinfo {title} {Nonadiabatic redshifts in high-order
  harmonic generation from solids},\ }\href
  {https://doi.org/10.1364/OE.25.023654} {\bibfield  {journal} {\bibinfo
  {journal} {Opt. Express}\ }\textbf {\bibinfo {volume} {25}},\ \bibinfo
  {pages} {23654} (\bibinfo {year} {2017})}\BibitemShut {NoStop}%
\bibitem [{\citenamefont {Ikemachi}\ \emph {et~al.}(2017)\citenamefont
  {Ikemachi}, \citenamefont {Shinohara}, \citenamefont {Sato}, \citenamefont
  {Yumoto}, \citenamefont {Kuwata-Gonokami},\ and\ \citenamefont
  {Ishikawa}}]{trajectory2017ikemachi}%
  \BibitemOpen
  \bibfield  {author} {\bibinfo {author} {\bibfnamefont {T.}~\bibnamefont
  {Ikemachi}}, \bibinfo {author} {\bibfnamefont {Y.}~\bibnamefont {Shinohara}},
  \bibinfo {author} {\bibfnamefont {T.}~\bibnamefont {Sato}}, \bibinfo {author}
  {\bibfnamefont {J.}~\bibnamefont {Yumoto}}, \bibinfo {author} {\bibfnamefont
  {M.}~\bibnamefont {Kuwata-Gonokami}},\ and\ \bibinfo {author} {\bibfnamefont
  {K.~L.}\ \bibnamefont {Ishikawa}},\ }\bibfield  {title} {\bibinfo {title}
  {Trajectory analysis of high-order-harmonic generation from periodic
  crystals},\ }\href {https://doi.org/10.1103/PhysRevA.95.043416} {\bibfield
  {journal} {\bibinfo  {journal} {Phys. Rev. A}\ }\textbf {\bibinfo {volume}
  {95}},\ \bibinfo {pages} {043416} (\bibinfo {year} {2017})}\BibitemShut
  {NoStop}%
\bibitem [{\citenamefont {Keldysh}(1965)}]{keldysh1965ionization}%
  \BibitemOpen
  \bibfield  {author} {\bibinfo {author} {\bibfnamefont {L.}~\bibnamefont
  {Keldysh}},\ }\bibfield  {title} {\bibinfo {title} {Ionization in the field
  of a strong electromagnetic wave},\ }\href@noop {} {\bibfield  {journal}
  {\bibinfo  {journal} {Soviet Physics JETP}\ }\textbf {\bibinfo {volume}
  {20}},\ \bibinfo {pages} {1307} (\bibinfo {year} {1965})}\BibitemShut
  {NoStop}%
\end{thebibliography}%

\end{document}